\documentclass[aps,pra,twocolumn,showpacs,superscriptaddress]{revtex4} 
\usepackage{graphicx}
\usepackage{amsmath}
\usepackage{amssymb}
\usepackage{epstopdf}
\usepackage{amsfonts}
\usepackage{dcolumn}
\usepackage{bigints}
\usepackage{xcolor}
\usepackage{hyperref}
\usepackage{comment}
\usepackage[normalem]{ulem}
\begin{document}
\title{Modeling the laser-pulse induced helium trimer dynamics}
\author{Q. Guan}
\address{Department of Physics and Astronomy, Washington State University, Pullman, Washington 99164-2814, USA}
\author{J. Kruse}
\address{Institut f\"ur Kernphysik, Goethe-Universit\"at, 60438 Frankfurt am Main, Germany}
\author{M. Kunitski}
\address{Institut f\"ur Kernphysik, Goethe-Universit\"at, 60438 Frankfurt am Main, Germany}
\author{R. D\"orner}
\address{Institut f\"ur Kernphysik, Goethe-Universit\"at, 60438 Frankfurt am Main, Germany}
\author{D. Blume}
\address{Homer L. Dodge Department of Physics and Astronomy,
  The University of Oklahoma,
  440 W. Brooks Street,
  Norman,
Oklahoma 73019, USA}
\address{Center for Quantum Research and Technology,
  The University of Oklahoma,
  440 W. Brooks Street,
  Norman,
Oklahoma 73019, USA}
\date{\today}

\begin{abstract}
Motivated by ongoing pump-probe spectroscopy experiments, this work develops a theoretical framework for describing the rovibrational wave packet dynamics that ensues when a single weakly-bound van der Waals trimer is exposed to a short, sub-picosecond linearly polarized pump laser pulse.
The 
intensity $I$ 
of the pump laser is chosen such that excitation and ionization of the electronic degrees of freedom are negligible while excitation of the wavepacket in the nuclear degrees of freedom is non-negligible.
The numerical treatment, which takes advantage of the fact that the laser 
pulse
is very short compared to typical molecular time scales, is based on a wave packet decomposition that utilizes hyperspherical coordinates. The framework is applied to the extremely floppy bosonic helium trimer. A convergence analysis 
of the partial wave decomposition is conducted. 
The kinetic energy release and orientation dynamics  are presented.
While the dynamics of more strongly-bound van der Waals trimers such as, e.g., the argon trimer display negligible coupling between vibrational and rotational degrees of freedom, rendering a description within a rigid-body picture appropriate, those of weakly-bound trimers display non-negligible coupling between vibrational and rotational degrees of freedom, rendering a description within a rigid-body picture inappropriate.
It is shown that a model that constructs the helium trimer dynamics from the dynamics of the helium dimer captures a number of key characteristics
of the alignment signal, including the interference between different angular momentum wave packet components.
\end{abstract}
\maketitle

\section{Introduction}

Weakly-bound van der Waals clusters have been studied extensively over the past several decades. There exist  many reasons why these systems are continuing to fascinate theorists and experimentalists alike. Bosonic helium clusters, e.g., are quantum liquids that exhibit microscopic superfluidity for around 60 or more atoms~\cite{PhysRevLett.50.1676,doi:10.1063/1.459525,doi:10.1126/science.279.5359.2083,https://doi.org/10.1002/anie.200300611,doi:10.1063/1.1310608}. Neon and argon clusters, in contrast, arrange in crystaline-like structures as the number of atoms increases~\cite{bressanini,doi:10.1063/1.460853,blume-greene-heavy}.
Since the cluster size can be varied in molecular beam experiments, these systems provide an ideal platform for studying the transition from 
the microscopic to the mesoscopic to the macroscopic scales. Moreover, large clusters serve as micro-laboratories. Argon clusters provide an ordered cage-like environment that has been utilized in matrix spectroscopy while helium clusters provide a unique  superfluid micro-laboratory~\cite{doi:10.1063/1.1359707,Stienkemeier_2006,doi:10.1063/1.1418746,doi:10.1146/annurev.physchem.49.1.1,argon-matrix1,argon-matrix2}. 

Small weakly-bound atomic helium and molecular para-hydrogen clusters  have attracted a great deal of attention due to their highly non-classical behavior, which can be quantified by the de Boer parameter~\cite{deboer1948a,deboer1948b,sevryuk}. Small bosonic helium clusters have enjoyed particularly broad attention since their interactions are characterized by a large positive two-body $s$-wave scattering length~\cite{doi:10.1063/1.438007,doi:10.1063/1.469978,doi:10.1063/1.460139,doi:10.1063/1.4712218,PhysRevLett.104.183003}. The large $s$-wave scattering length implies that a subset of the system's properties can be accurately captured by effective low-energy theories that replace the true atom-atom interaction potential by simpler approximate model potentials that are designed to reproduce the $s$-wave scattering length and two-body binding
 energy~\cite{BRAATEN2006259,BRAATEN2007120,Naidon_2017,greene-rev-mod-physics}.
The large scattering length places small helium clusters inside the universal or near-universal window, where systems of vastly different sizes, such as the helium trimer and the triton, share certain characteristics, provided their energies and lengths are scaled appropriately~\cite{BRAATEN2006259,BRAATEN2007120,Naidon_2017,greene-rev-mod-physics}. The link between these systems is established through Efimov's zero-range theory~\cite{efimov70}.

While the literature on static properties of weakly-bound van der Waals clusters is vast,
dynamic studies 
have so far focused on electronically excited states~\cite{dynamic_electronic1,dynamic_electronic2,dynamic_electronic3}.
Dynamic studies of van der Waals trimers in their electronic ground state are just starting to emerge~\cite{QiWei,friedrich1998,nielsen1999,bruch2000,blume-guan,nature-physics}. Specifically,
it was recently demonstrated that state-of-the-art experiments~\cite{nature-physics,our-experiment-theory-paper} have the capability to (i)  prepare  isolated helium dimers and trimers, (ii)  apply a pump pulse that operates at the ``sweet spot'' where dynamics of the nuclear but not of the electronic degrees of freedom is triggered, and (iii) image the atomic positions as a function of the delay between the pump and probe pulses using a COLTRIMS reaction microscope~\cite{coltrims,Ulrich2011,PhysRevA.98.050701}. 
Unlike in heavier and more strongly-bound molecules, where pump-probe spectroscopy has been used extensively to study rotational revivals~\cite{RMP_alignment,review_lemeshko}---with applications in areas as diverse as high-harmonic generation~\cite{RMP_alignment},
generation of Bloch oscillations~\cite{floss2015},
and probe of fast collisional dissipative processes~\cite{lin2020}---, in light van der Waals trimers, the internal motion cannot be reduced to that of a rigid body. Instead, the rotational degrees of freedom  couple to the vibrational degrees of freedom, resulting in unique and essentially unexplored rovibrational dynamics of a system whose initial state is, at least approximately, captured by a universal theory. 
The advances on the experimental side motivate the development of a theoretical three-body framework that accounts for the full dynamics of the nuclear wave packet.

The remainder of this paper is organized as follows. Section~\ref{sec_theory} introduces the system under study and the wave packet decomposition. 
Section~\ref{sec_numerics} discusses  aspects of our numerical implementation and the convergence of the wave packet decomposition.
Section~\ref{sec_observables}   motivates and defines several observables.
Section~\ref{sec_results} analyzes the wave packet dynamics of the laser-kicked helium trimer, with particular emphasis on characterizing the bound and scattering portions of the wave packet as well as the alignment and dynamics associated with the Euler angles.
Finally, Sec.~\ref{sec_conclusion} summarizes and provides an outlook.
The appendices introduce simple models that aid in developing physical intuition of the trimer dynamics.
Appendix~\ref{appendix_rigid} derives the dynamics expected for a rigid triatomic molecule while 
Appendix~\ref{appendix_dimer} presents 
a simple model for the helium trimer dynamics that utilizes only input from a simulation of the dimer.

\section{Theoretical framework}
\label{sec_theory}

\subsection{System under study}
\label{sec_theory_ham}

The system Hamiltonian $H_{\text{tot}}(t)$ consists of the time-independent Hamiltonian $H_{\text{trimer}}$ that describes the van der Waals trimer, which is composed of three identical bosons, and the time-dependent  Hamiltonian $H_{\text{trimer-laser}}(t)$ that describes the interaction of the trimer with the laser,
\begin{eqnarray}
    H_{\text{tot}}(t)=
    H_{\text{trimer}}+H_{\text{trimer-laser}}(t).
\end{eqnarray}
The trimer Hamiltonian  contains the kinetic energy operator and the pairwise
atom-atom interaction,
\begin{eqnarray}
H_{\text{trimer}}=
\sum_{j=1}^3
\frac{-\hbar^2}{2 \mu} \frac{\partial^2}{\partial \vec{r}_j^2}
+ V_{\text{trimer}}(r_{12},r_{13},r_{23}),
\end{eqnarray}
where
\begin{eqnarray}
  V_{\text{trimer}}(r_{12},r_{13},r_{23})
  =
  \sum_{j=1}^2 \sum_{k=j+1}^3
  V_{\text{aa}}(r_{jk}).
  \end{eqnarray}
Here, $\mu$ denotes the atom  mass,
$\vec{r}_j$ the position vector of the $j$th atom, and
$r_{jk}$ the interatomic distance between atom $j$ and atom $k$,
$\vec{r}_{jk}=\vec{r}_j-\vec{r}_k$ and $r_{jk}=|\vec{r}_{jk}|$.
The atom-atom Born-Oppenheimer potential $V_{\text{aa}}(r_{jk})$
accounts for the short-distance
repulsion due to the overlapping electron clouds and the large-distance dispersive 
van der Waals attraction. 
Our calculations take advantage of the fact that $H_{\text{tot}}(t)$ separates into the relative and center-of-mass degrees of freedom,
    $H_{\text{tot}}(t)=H_{\text{rel}}(t)+
    T_{\text{cm}}$,
where $T_{\text{cm}}$ denotes the kinetic energy operator 
of the center-of-mass degrees of freedom~\cite{footnote_cm}.
The relative part $H_{\text{trimer,rel}}$ of $H_{\text{trimer}}$ is defined through 
\begin{eqnarray}
    H_{\text{trimer,rel}}=H_{\text{trimer}}-T_{\text{cm}}.
    \end{eqnarray}

We are interested in van der Waals trimers that support at least one $J^{\Pi}=0^+$ bound state, where $J$ denotes the relative orbital angular momentum quantum number and $\Pi$ the relative parity of the trimer.
 Using Whitten-Smith's coordinates~\cite{whitten-smith2,johnson1980,kendrick1990}, namely the hyperradius 
 $\rho$  and the hyperangles $\theta$ and $\phi$, we denote
the ground  state of $H_{\text{trimer,rel}}$ by $\psi_{\text{trimer}}^{(\text{ground})}(\rho, \theta,\phi)$.
The ranges of the coordinates are $\rho \in [0,\infty]$, $\theta \in [0,\pi/4]$,
and $\phi \in [0,\pi/3]$~\cite{whitten-smith2,johnson1980,kendrick1990,parker-hyper}.
Here, we only provide an explicit expression for the hyperradius $\rho$,
\begin{eqnarray}
    \rho^2 = \frac{1}{\sqrt{3}}
    \left( r_{12}^2 + r_{13}^2 + r_{23}^2 \right).
\end{eqnarray}
Expressions for $\theta$ and $\phi$ can be obtained by  inverting the  expressions for the internuclear distances $r_{jk}$~\cite{suno2002}:
\begin{eqnarray}
\label{eq_r12new}
r_{12} = \frac{\rho}{3^{1/4}}
\left[
1+ \cos(2 \theta) \cos(2 \phi) 
\right]^{1/2},
\end{eqnarray}
\begin{eqnarray}
\label{eq_r13new}
r_{13} = \frac{\rho}{3^{1/4}}
\left[
1+ \cos(2 \theta) \cos \left(2 \phi - \frac{2 \pi}{3} \right)
\right]^{1/2},
\end{eqnarray}
and
\begin{eqnarray}
\label{eq_r23new}
r_{23} = \frac{\rho}{3^{1/4}}
\left[
1+ \cos(2 \theta) \cos \left(2 \phi 
+ \frac{2 \pi}{3} \right) \right]^{1/2}.
\end{eqnarray}
Owing to the bosonic exchange symmetry,
the trimer ground state wave function obeys the 
boundary conditions~\cite{suno2002,suno2008}
\begin{eqnarray}
\frac{\partial \psi_{\text{trimer}}^{(\text{ground})}(\rho, \theta,\phi)}{\partial \phi} \Bigg| _{\phi=0}=
\frac{\partial \psi_{\text{trimer}}^{(\text{ground})}(\rho, \theta,\phi)}{\partial \phi} \Bigg| _{\phi=\pi/3}=0.\nonumber \\
\end{eqnarray}
We note that the
solutions for $\phi \in [\pi/6,\pi/3]$ can be obtained from those
for  $\phi \in [0,\pi/6]$ by applying a reflection operation~\cite{doi:10.1063/1.482027}.
Our approach, however, uses the ``enlarged'' $\phi$-range throughout.
We employ the normalization 
\begin{eqnarray}
    \int_0^{\infty} 
    \int_0^{\frac{\pi}{3}} \int_0^{\frac{\pi}{4}} |\psi_{\text{trimer}}^{(\text{ground})}
  (\rho,\theta,\phi)
    |^2 
    \frac{\sin(4 \theta) }{4}  \rho^5 d \theta d \phi d \rho=\nonumber \\ 1;\nonumber \\ 
    \end{eqnarray}
    for later convenience, we introduce the volume element
$d V_{\text{hyper}}$,
\begin{eqnarray}
  dV_{\text{hyper}}= 
  \frac{1}{4}  \sin(4 \theta) \rho^5 d \theta d \phi d \rho.
\end{eqnarray}
As discussed in more detail in Sec.~\ref{sec_theory_decomposition}, $\psi_{\text{trimer}}^{(\text{ground})}
  (\rho,\theta,\phi)$ serves as the initial state for our time-dependent studies.

The laser-trimer interaction is modeled    
assuming that the laser interacts with each of the atom pairs of the trimer as if the pairs were independent, i.e., the laser-trimer interaction 
is written as
\begin{eqnarray}
V_{\text{laser-trimer}}(r_{12},\vartheta_{12}, r_{13} ,\vartheta_{13}, r_{23},\vartheta_{23},t)
= 
\nonumber \\
-\frac{1}{2} \epsilon^2(t) \sum_{j=1}^2 \sum_{k>j}^3
V_{\text{laser-dimer}}(r_{jk},\vartheta_{jk}),
\end{eqnarray}
where
\begin{eqnarray}
\epsilon^2(t)=
|\overline{\epsilon}|^2 \exp \left[ -4 \, \text{ln} 2 \left(\frac{t}{\tau} \right)^2 \right]
\end{eqnarray}
and~\cite{buckingham1973,cencek2011}
\begin{eqnarray}
\label{eq_laser_dimer}
V_{\text{laser-dimer}}(r_{jk},\vartheta_{jk})
=
\alpha_{\text{int}}(r_{jk})
+ 
\nonumber \\
\frac{1}{3} \sqrt{\frac{16 \pi}{5}} \beta_{\text{int}}(r_{jk} )
Y_{2,0}(\vartheta_{jk}).
\end{eqnarray}
The Gaussian laser pulse is written in terms of the pulse duration  or  full-width-half-maximum (FWHM) $\tau$
and the mean electric field strength
$\overline{\epsilon}$, which is obtained by averaging over the fast oscillating electromagnetic field~\cite{friedrich1995}.
In Eq.~(\ref{eq_laser_dimer}), $Y_{2,0}(\vartheta_{jk})$ denotes the spherical harmonic and
 $\vartheta_{jk}$
is defined as the angle between the space-fixed
$z$-axis and the 
internuclear distance vector $\vec{r}_{jk}$ in the laboratory frame.
The quantities $\alpha_{\text{int}}(r_{jk})$ and
 $\beta_{\text{int}}(r_{jk})$ denote, respectively, the distance-dependent isotropic and anisotropic polarizabilities.
 The laser-molecule interaction is applicable to linearly polarized light whose polarization vector lies along the space-fixed $z$-axis. It is assumed that the laser intensity is chosen such that the laser does not lead to ionization or electronic excitations~\cite{nature-physics,becker,ADK,Voigtsberger2014,doi:10.1126/science.aaa5601,Zeller14651}.
 
\begin{figure}
    \includegraphics[width=0.4\textwidth]{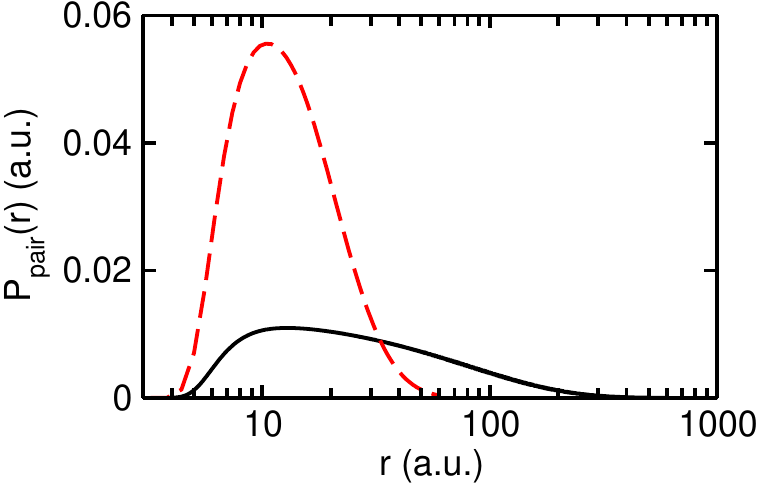}
    \caption{Solid and dashed lines show the pair distribution function $P_{\text{pair}}(r)$ of the helium ground state dimer and the helium ground state trimer.
     The normalization is chosen such that
    $ \int_0^{\infty} P_{\text{pair}}(r)dr=1$.
    The $x$- and $y$-axes both employ atomic units (a.u.); specifically, the $x$- and $y$-axes use $a_0$ and $(a_0)^{-1}$, respectively.}
    \label{fig_pair}
\end{figure}

 Section~\ref{sec_results} considers three identical $^4$He atoms.
We use $\mu=7296.30$~m$_e$ (m$_e$ denotes the electron mass) 
and the helium-helium potential from Ref.~\cite{doi:10.1063/1.4712218},
which yields a dimer energy of $E_{\text{dimer}}^{(\text{ground})}=-1.62$~mK and
a dimer scattering length of $a_{\text{dimer}}=90.4$~\AA~\cite{doi:10.1063/1.4712218,PhysRevLett.104.183003}.
The trimer 
ground state energy is 
$E_{\text{trimer}}^{(\text{ground})}=-131.8$~mK~\cite{PhysRevA.85.062505,AJ}.
While the helium dimer does not support a bound excited state, the helium trimer supports
a single bound excited state with energy $E_{\text{trimer}}^{(\text{excited})}=-2.65$~mK in the $J=0$ subspace~\cite{PhysRevA.85.062505} (the $J>0$ subspaces do not support any three-body bound states~\cite{esryJPB}).
The helium trimer excited bound state is an Efimov state~\cite{PhysRevLett.38.341,doi:10.1063/1.450912,Nielsen_1998,Blume2015,doi:10.1126/science.aaa5601,Kolganova2011,PhysRevA.54.394}.
We denote the bound excited trimer wave function 
by
$\psi_{\text{trimer}}^{(\text{excited})}(\rho, \theta,\phi)$. The fact that the trimer ground state binding energy is notably larger than that of the dimer suggests that the trimer ground state is much more compact than the dimer ground state. Indeed, the average interparticle spacing of the trimer is about seven times smaller than that of the dimer~\cite{doi:10.1126/science.aaa5601}. Inspection of the pair distribution functions (see Fig.~\ref{fig_pair}) shows that the trimer ground state is, despite of its smaller size, a rather delocalized diffuse quantum object.

 Earlier work has shown that the laser-pulse induced helium dimer dynamics is captured accurately by the 
laser-molecule interaction $V_{\text{laser-dimer}}(r_{12},\vartheta_{12})$~\cite{nature-physics}, with polarizabilities taken from  Ref.~\cite{cencek2011}. 
Even though the three-body contributions to the polarizability has been calculated for 
helium~\cite{polarizability-three-body1,polarizability-three-body2,polarizability-three-body3}, they 
are not included in the polarizability model introduced above. Reference~\cite{polarizability-three-body1} shows that intrinsic three-body contributions for helium are, in the parameter regime investigated in the present work, roughly two orders of magnitude smaller than the two-body contributions. The applicability of the additive pair-polarization model is further confirmed by the fact that results obtained using this model compare favorably with experimental data~\cite{our-experiment-theory-paper}.

\subsection{Wave packet decomposition}
\label{sec_theory_decomposition}

Our goal is to
determine the time-dependent wave packet $\Psi(\rho,\theta,\phi,\beta,\gamma,t)$ that ensues 
due to the interaction of the van der Waals trimer with the  external  laser. Here, the 
Euler angles $\beta$ and $\gamma$ describe the orientation of the trimer
in space (the dynamics is independent of the Euler angle $\alpha$).
Our definition of the Euler angles follows Goldstein and Poole~\cite{goldstein-poole},
with $\gamma$, $\beta$, and $\alpha$ denoting rotations about the body-fixed
$z$-axis, nodal line, and
new body-fixed $z$-axis, respectively.
Our primary interest lies in 
determining the dynamics after the laser  pulse  intensity has decayed to a negligibly small value on time scales that are larger than $\tau$.
The FWHM $\tau$ is chosen to be  short compared to  the ``characteristic times'' of the trimer such as the time scale(s) that derive(s) from the energy level spacing(s) of the bound trimer eigen states.

To solve the time-dependent Schr\"odinger equation 
\begin{eqnarray}
\imath \hbar \frac{\partial \Psi(\rho,\theta,\phi,\beta,\gamma,t)}{\partial t} = 
H_{\text{rel}}(t) 
\Psi(\rho,\theta,\phi,\beta,\gamma,t),
\end{eqnarray}
which is, for finite laser intensity, a six-dimensional (5+1 dimensions) partial differential equation, we employ a partial wave decomposition.
The fact that the time evolution is independent of $\alpha$
can be seen by rewriting $\vartheta_{12}$, $\vartheta_{13}$, and $\vartheta_{23}$
in terms of $\rho$, $\theta$, $\phi$, $\beta$, and $\gamma$ (see Appendix~\ref{appendix_rigid}).
Our initial state is chosen to be the trimer ground state:
\begin{eqnarray}
  \label{wave_initial}\Psi(\rho,\theta,\phi,\beta,\gamma,0) = \psi_{\text{trimer}}^{(\text{ground})}
  (\rho,\theta,\phi)  \sqrt{\frac{1}{8 \pi^2}}.
\end{eqnarray}
The factor of $(8 \pi^2)^{-1/2}$ is added to ensure normalization over the entire space,
\begin{eqnarray}
  \int  |\Psi(\rho,\theta,\phi,\beta,\gamma,0)|^2 dV_{\text{hyper}} dV_{\text{euler}} = 1,
\end{eqnarray}
where the volume element $dV_{\text{euler}}$ reads
\begin{eqnarray}
  dV_{\text{euler}} = \sin(\beta) d \alpha d \beta d \gamma
\end{eqnarray}
with $\alpha \in [0,2 \pi]$, $\beta \in [0,\pi]$, and $\gamma \in [0, 2\pi]$.
The chosen initial state applies to an experiment where the population of the excited states prior to the turning-on of the pump laser pulse is negligibly small.

 The pump 
laser couples different $J$ subspaces, thereby introducing an explicit  dependence of the time-dependent wave packet on
the Euler angles. After the laser pulse has decayed to essentially zero, different $J$ channels are decoupled, i.e., $J$ is a good quantum number before the
pump laser is turned on and after the pump laser is turned off. 
Since treating the dynamics 
in the presence of the laser pulse is more challenging numerically than treating the dynamics in the absence of the laser pulse, we employ a simplified  approximate
approach that takes advantage of the fact that $\tau$ is small.  Specifically, we approximate the
time-dependent laser-dimer potential by a $\delta$-function potential
whose strength is adjusted such that it coincides with
the area under the Gaussian pulse, i.e., we set~\cite{nature-physics}
\begin{eqnarray}
\epsilon^2(t) = | \overline{\epsilon}|^2 \sqrt{\frac{\pi \tau^2}{4 \, \ln 2}} \delta(t).
\end{eqnarray}
Using this $\delta$-function approximation,
the trimer wave packet at time $t=0^+$ reads
\begin{eqnarray}
\label{eq_after_kick}
\Psi(\rho,\theta,\phi,\beta,\gamma,0^+)= 
\nonumber \\
\exp \left[ \imath \overline{\varphi}(\rho,\theta,\phi,\beta,\gamma,0^+) \right]
\Psi(\rho,\theta,\phi,\beta,\gamma,0),
\end{eqnarray}
where the spatially-dependent 
phase factor
$\overline{\varphi}(\rho,\theta,\phi,\beta,\gamma,0^+)$ is given by
\begin{eqnarray}
\label{eq_phase2}
\overline{\varphi}(\rho,\theta,\phi,\beta,\gamma,0^+)
=  
C
\sum_{j=1}^2\sum_{k>j}^3
V_{\text{laser-dimer}}(r_{jk},\vartheta_{jk}) \nonumber \\
\end{eqnarray}
with
\begin{eqnarray}
  \label{eq_c}
C = \frac{1}{2} |\overline{\epsilon}|^2
\sqrt{\frac{\pi \tau^2}{4 \, \ln2}} \frac{1}{\hbar}.
  \end{eqnarray}
Recall,
$\vartheta_{jk}$
denotes  the angle between the polarization vector of the laser and the 
internuclear distance vector $\vec{r}_{jk}$.
Since $\exp [ \imath \overline{\varphi}(\rho,\theta,\phi,\beta,\gamma,0^+)]$
is ``just'' a phase factor [implying $|\exp(\imath \overline{\varphi})|^2=1$], the normalization of $\Psi(\rho,\theta,\phi,\beta,\gamma,0)$ implies that the wave packet at $0^+$ is also normalized.
The phase $\overline{\varphi}(\rho,\theta,\phi,\beta,\gamma,0^+)$ approaches zero in the large $\rho$ limit. Its dependence on the hyperspherical coordinates is discussed in Sec.~\ref{sec_numerics_details}.

Importantly, because  the laser pulse contains a $\delta$-function in time (i.e., because it has decayed to zero at $t=0^+$), the remaining task is
to time evolve the wave packet
$\Psi(\rho,\theta,\phi,\beta,\gamma,0^+)$
under the Hamiltonian $H_{\text{trimer,rel}}$. We refer the reader to Ref.~\cite{nature-physics} for a study that benchmarks the delta-function laser-molecule model in the context of the helium-dimer dynamics. Treatment of the more sophisticated laser-trimer interaction
model, which explicitly accounts for the Gaussian envelope of the laser intensity, is
relegated to future work.

To perform the time evolution, we take advantage of the fact that $J$
is---for the delta-function model---a good quantum number for $t>0^+$.
Specifically, we decompose the wave packet
$\Psi(\rho,\theta,\phi,\beta,\gamma,t)$
into partial wave components $\Psi^{(J)}(\rho,\theta,\phi,\beta,\gamma,t)$,
\begin{eqnarray}
  \label{eq_wave_j}
  \Psi(\rho,\theta,\phi,\beta,\gamma,t)=
  \sum_{J=0,2,\cdots}
\Psi^{(J)} (\rho,\theta,\phi,\beta,\gamma,t),
\end{eqnarray}
which are written in terms of angle-dependent but time-independent 
channel functions $\Phi^{(J)}_{m,n}(\theta,\phi,\beta,\gamma)$ and $\rho$- and $t$-dependent hyperradial weights
$F^{(J)}_{m,n}(\rho,t)$,
\begin{eqnarray}
  \label{eq_expand}
\Psi^{(J)} (\rho,\theta,\phi,\beta,\gamma,t)=
\nonumber \\
    \sum_{m=0,\pm 6,\dots} \sum_{n=0,1,\cdots} F^{(J)}_{m,n}(\rho,t)\Phi^{(J)}_{m,n}(\theta,\phi, \beta,\gamma).
\end{eqnarray}
For isotropic  initial states with $J^{\Pi}=0^+$, the functional form of the laser-trimer interaction only populates even $J$ channels with positive relative parity $\Pi$. 

The channel functions $\Phi^{(J)}_{m,n}(\theta,\phi, \beta,\gamma)$
are expanded in terms of 
a complete set of functions in $\phi$,
a complete set of functions in $\theta$,
and a complete set of functions in $\beta$ and $\gamma$,
\begin{eqnarray}
\label{eq_channel}
\Phi^{(J)}_{m,n}(\theta,\phi, \beta,\gamma)=
\nonumber \\ \sum_{M=-J,-J+2,\cdots,J} 
    \sqrt{\frac{3}{\pi}}\exp( \imath m \phi)
    P^{(J,m)}_{M,n}(\theta) \times \nonumber \\
    \sqrt{\frac{2J+1}{8 \pi^2}}
    D_{0,M}^{(J)}(0,\beta,\gamma),
\end{eqnarray}
where $m$ takes the values $0,\pm 6,\pm 12,\cdots$~\cite{whitten-smith2}.
The factor of $\sqrt{3/\pi}$ ensures that the functions
in $\phi$ are normalized:
\begin{eqnarray}
\label{orthonormal1}
\frac{3}{\pi}  \int_0^{\pi/3} \exp( -\imath m' \phi)
  \exp( \imath m \phi)
  d \phi = \delta_{m',m}.
  \end{eqnarray}
In Eq.~(\ref{eq_channel}), $D_{M',M}^{(J)}(\alpha,\beta,\gamma)$
denotes the Wigner-D function~\cite{zare-book},
\begin{eqnarray}
\label{orthonormal2}
  \int
      [D_{M''',M'}^{(J')}(\alpha,\beta,\gamma)]^*
      D_{M'',M}^{(J)}(\alpha,\beta,\gamma) dV_{\text{euler}}
      = \nonumber \\
      \frac{8 \pi^2}{2J+1} \delta_{J',J}\delta_{M',M} \delta_{M''',M''}
      .
\end{eqnarray}
The Hilbert space in the $\theta$ coordinate is spanned by the functions
$P^{(J,m)}_{M,n}(\theta)$,
which are constructed---for each allowed $(J,m)$ combination---by solving the set of
coupled one-dimensional differential equations
in
the $\theta$ degree of freedom numerically,
\begin{widetext}
\begin{eqnarray}
\label{eq_ptheta_main}
\Bigg[
\frac{4}{\tan (4 \theta)} 
\frac{\partial}{\partial \theta}
+  
\frac{\partial^2}{\partial \theta^2}
-
\frac{m^2}{\cos^2( 2 \theta)}
-
\frac{2 J(J+1) -2 M^2}{\sin^2(2 \theta)}
-
\frac{M^2}{ \cos^2( 2 \theta)}
-
\frac{2 m M \sin(2\theta) }{ \cos^2( 2 \theta)}
\Bigg] {P}^{(J,m)}_{M,n}(\theta)
\nonumber \\ 
-
\frac{\cos(2 \theta) A^+_{J,M}}{\sin^2(2 \theta)}
 {P}^{(J,m)}_{M+2,n}(\theta) 
-
\frac{\cos(2 \theta) A^-_{J,M}}{\sin^2(2 \theta)}
 {P}^{(J,m)}_{M-2,n}(\theta) 
 = 
\nonumber \\
\overline{E}^{(J,m)}_{n} {P}^{(J,m)}_{M,n}(\theta) ,
\end{eqnarray}
\end{widetext}
  where  
\begin{eqnarray}
\label{eq_aplusminus}
A^{\pm}_{J,M}=\sqrt{J(J+1)-M(M \pm 1)}  \times \nonumber \\
\sqrt{J(J+1)-(M \pm 1)(M \pm 2)}. 
\end{eqnarray}
Since
$M$ can take the values $-J,-J+2,\cdots,J$,
the number of coupled equations is $J+1$.
For each $(J,m)$ combination,
the orthonormality condition reads 
\begin{eqnarray}
\label{orthonormal3}
  \sum_{M=-J,\cdots,J} 
  \int_0^{\pi/4} [P^{(J,m)}_{M,n'}(\theta)]^*
      P^{(J,m)}_{M,n}(\theta)
      \frac{\sin(4 \theta)}{4}  d \theta = \nonumber \\ \delta_{n,n'}.
  \end{eqnarray}
The subscript $n$ ($n=0,1,\cdots$) enumerates the
solutions to the coupled equations with dimensionless eigen value $\overline{E}_{n}^{(J,m)}$.
Equations~\eqref{orthonormal1}-\eqref{orthonormal3} imply orthonormality of the channel functions,
\begin{widetext}
\begin{eqnarray}
\label{orthonormal4}
\int \left[\Phi^{(J')}_{m',n'}(\theta,\phi, \beta,\gamma)\right]^*\Phi^{(J)}_{m,n}(\theta,\phi, \beta,\gamma)\frac{1}{4}\sin(4\theta)d\theta d\phi dV_{\text{euler}}=\delta_{J',J}\delta_{m',m}\delta_{n',n}.
\end{eqnarray}
The hyperradial weights 
$F^{(J)}_{m,n}(\rho,t)$
at $t=0^+$
are given by
\begin{eqnarray}
 F^{(J)}_{m,n}(\rho,0^+)=  
  \int \left[\Phi^{(J)}_{m,n}(\theta,\phi,\beta,\gamma)\right]^* \Psi(\rho,\theta,\phi,\beta,\gamma,0^+)
  \frac{1}{4} \sin(4 \theta) d \theta d\phi    
  dV_{\text{euler}}. 
  \end{eqnarray}
  \end{widetext}
The time evolution of the hyperradial weights is obtained by solving a (1+1)-dimensional set
of second-order differential equations:
\begin{eqnarray}
  \imath \hbar
  \frac{\partial F^{(J)}_{m,n}(\rho,t)}{\partial t}
  = 
  \frac{-\hbar^2}{2M_{\text{hyper}}\rho^5}
    \frac{\partial}{\partial \rho}\rho^5\frac{\partial}{\partial \rho}  F^{(J)}_{m,n}(\rho,t)
    - \nonumber \\
    \frac{\hbar^2}{2 M_{\text{hyper}}\rho^2} \overline{E}_{n}^{(J,m)} F_{m,n}^{(J)}(\rho,t) + \nonumber \\
    \sum_{m'=0,\pm 6,\cdots} \sum_{n'=0,1,\cdots}
      W^{(J)}_{m,n;m',n'}(\rho)F^{(J)}_{m',n'}(\rho,t),
\end{eqnarray}
where $M_{\text{hyper}}$ denotes the hyperradial mass, $M_{\text{hyper}}=3^{-1/2} \mu$,
and $W^{(J)}_{m,n;m',n'}(\rho)$ the coupling matrix elements
that account for 
the interaction potential $V_{\text{trimer}}(r_{12},r_{13},r_{23})$,
\begin{widetext}
\begin{eqnarray}
    W^{(J)}_{m,n;m',n'}(\rho)= 
    \int
    \left[\Phi^{(J)}_{m,n}(\theta,\phi,\beta,\gamma)\right]^* V_{\text{trimer}}(r_{12},r_{13},r_{23}) 
    \Phi^{(J)}_{m',n'}(\theta,\phi,\beta,\gamma)
  \frac{1}{4} \sin(4 \theta) d \theta d\phi    
  dV_{\text{euler}}
    ,
\end{eqnarray}
\end{widetext}
where the internuclear distances 
$r_{jk}$ need to be expressed in terms of hyperspherical coordinates [see Eqs.~(\ref{eq_r12new})-(\ref{eq_r23new})].

We note in passing that the approach developed in this section can be quite straightforwardly generalized to non-identical three-particle systems as well as three-particle systems that contain identical fermions~\cite{Naidon_2017,blume-review-article,D_Incao_2018}.
The next section discusses aspects of our numerical implementation for three identical bosons.

\section{Implementation details and convergence}
\label{sec_numerics}
While the outlined framework can be applied to a range of weakly-bound bosonic three-atom clusters, throughout this section we consider the three-helium-atom cluster as an example. 

\subsection{Numerical grid}
\label{sec_numerics_details}

The bound eigen states and associated eigen energies of $H_{\text{trimer,rel}}$ are obtained by calculating the fixed-$\rho$ adiabatic eigen values and eigen states using a B-spline basis in $\theta$ and $\phi$ and by subsequently solving a set of coupled equations in $\rho$~\cite{doi:10.1063/1.482027,esry1999}.
While the eigen state of the excited Efimov trimer is not needed to perform the time propagation, the projection of the $0^+$-wave packet onto the Efimov state yields valuable physical insight. 
For the hyperangular degrees of freedom, we use non-linear grids that vary with the hyperradius; we use $140$ and 
$240$
grid points in $\theta$ and $\phi$, respectively~\cite{doi:10.1063/1.482027}.
The grid in the hyperradius $\rho$ is also non-linear; it is chosen such that the amplitude of the ground state wave function at the largest $\rho$ value is
negligibly small. 
We use 900 points between $\rho=5$~a$_0$ and $1,500$~a$_0$ for the initial state preparation and time evolution.

\begin{figure}
\vspace*{-0.2in}
\includegraphics[width=0.5\textwidth]{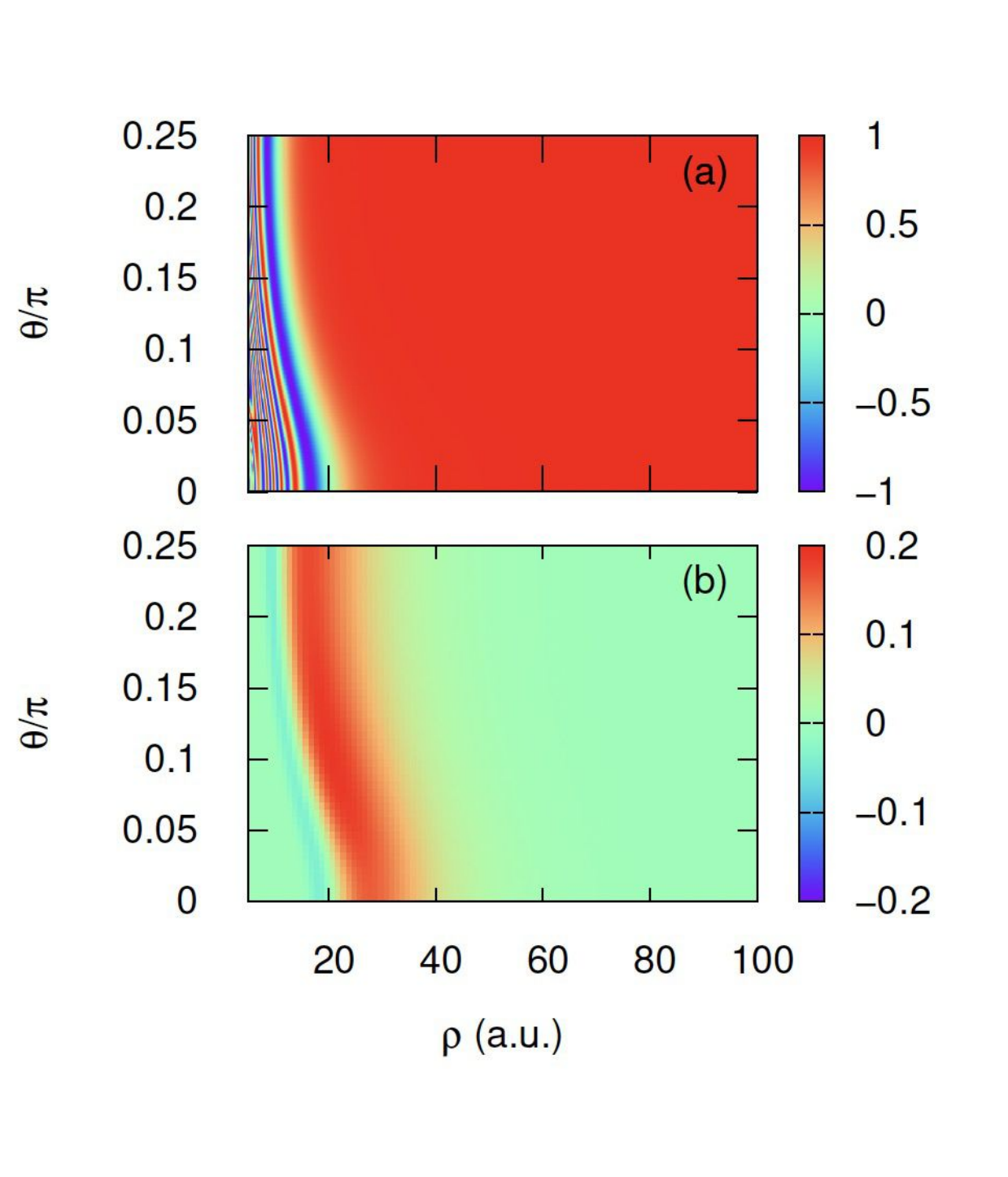}
\vspace*{-0.6in}
    \caption{The real part of (a) $\exp[\imath\overline{\varphi}(\rho,\theta,\phi,\beta,\gamma,0^+)]$ and  (b) $\exp[\imath \overline{\varphi}(\rho,\theta,\phi,\beta,\gamma,0^+)]|\psi^{(\text{ground})}_{\text{trimer}}(\rho,\theta,\phi)|^2\rho^{5}$ for  $\beta=0$, $\gamma=0$, and $\phi=\pi/12$. 
    The legends on the right color code the dimensionless quantities
    $\mbox{Re}\{\exp[\imath\overline{\varphi}(\rho,\theta,\phi,\beta,\gamma,0^+)]\}$ and $\mbox{Re}\{\exp[\imath \overline{\varphi}(\rho,\theta,\phi,\beta,\gamma,0^+)]|\psi^{(\text{ground})}_{\text{trimer}}(\rho,\theta,\phi)|^2\rho^{5}\}$.
    A $\delta$-function pulse corresponding to $I=3.5\times 10^{14}$~W/cm$^2$ and $\tau=331$~fs ($|\overline{\epsilon}|^2=0.9972\times10^{-2}$~a.u.) is used.
 }
\label{fig_phase}
\end{figure}

The wave packet for $t=0^+$ depends on the phase $\overline{\varphi}(\rho,\theta,\phi,\beta,\gamma,0^+)$.
Using a delta-function pulse corresponding to $I=3.5\times 10^{14}$~W/cm$^2$ and $\tau=331$~fs, Fig.~\ref{fig_phase}(a) shows the  real part of $\exp[\imath \overline{\varphi}(\rho,\theta,\phi,\beta,\gamma,0^+)]$
as functions of $\theta$ and $\rho$ for $\phi=\pi/12$, $\beta=0$, and $\gamma=0$. Since the phase approaches zero for large $\rho$, the real part of the exponential approaches one. For smaller $\rho$ ($\rho \lesssim 20$~a.u.), the real part of the exponential displays highly oscillatory behavior. 
To understand the relevance of these oscillations, Fig.~\ref{fig_phase}(b) multiplies the real part of $\exp[\imath \overline{\varphi}(\rho,\theta,\phi,\beta,\gamma,0^+)]$
by the scaled ground state density
$|\psi^{(\text{ground})}_{\text{trimer}}(\rho,\theta,\phi)|^2\rho^{5}$.
It can be seen that the small amplitude of the ground state density at small hyperradii strongly suppresses the oscillations at  $\rho \lesssim 10-15$~a.u.. For the parameters chosen, the real part of $\exp[\imath \overline{\varphi}(\rho,\theta,\phi,\beta,\gamma,0^+)]|\psi^{(\text{ground})}_{\text{trimer}}(\rho,\theta,\phi)|^2\rho^{5}$ changes---starting at large $\rho$---from positive to negative to positive over the range of $\rho$ shown (additional oscillations at smaller $\rho$ exist but are not visible on the scale shown).  Figure~\ref{fig_phase} suggests that the phase $\overline{\varphi}(\rho,\theta,\phi,\beta,\gamma,0^+)$ can be treated perturbatively for a large, but not the entire, portion of the configuration space. A similar conclusion can be drawn for other $\theta$, $\beta$, and $\gamma$ values.

The integrals that determine the radial weights $F^{(J)}_{m,n}(\rho, 0^+)$ and the coupling matrix elements $W^{(J)}_{m,n;m',n'}(\rho)$ are evaluated numerically using linear grids in $\theta$ and $\phi$.
The initial state on the linear hyperangular grid is obtained through interpolation of  
the ground state, which is determined on non-linear grids (see above). 
Using successively more grid points in $\theta$, we checked that $500$ points in $\theta$ yield converged results for $\overline{E}_n^{(J,m)}$ and
$P_{M,n}^{(J,m)}(\theta)$ for all $J$, $m$, and $n$ considered.
Similarly, using successively more grid points in $\phi$, we checked that $240$ points in $\phi$ yield converged results for $W^{(J)}_{m,n;m',n'}(\rho)$ and $F^{(J)}_{m,n}(\rho, 0^+)$ for all $m$, $n$, $m'$, $n'$, and $\rho$ considered. 
We checked that  unphysical reflections during the wave packet propagation at the largest $\rho$ value covered by the initial ground state (i.e., the edge of the box) 
have a negligible effect on the dynamics.

Starting with the
$t=0^+$ weights, we obtain the time evolution by expanding the
propagator in terms of Chebychev polynomials~\cite{kosloff}. The derivatives in $\rho$ are calculated via finite-differencing, accounting for the non-linear spacing of the grid points.
The numerical effort depends on the chosen time step and the order of the highest Chebychev polynomial. We  work with time steps around $2$~a.u. and  Chebychev polynomials of order up to $30$.
If a larger/smaller time step is used, the order of the highest Chebychev polynomial can be increased/decreased to reach comparable accuracy. 
Time step convergence is checked by making sure that the wave packet remains normalized at all times and by making sure that the time-evolved wave packet is essentially independent of the time step used to do the wave packet propagation.

During the time evolution (the maximum time considered in this work is $50$~ps), the total probability of the wave packet [i.e., the integral of $|\Psi(\rho,\theta,\phi,\beta,\gamma,t)|^2$ over all spatial coordinates] changes by less than $10^{-3}$~\% 
(while this value depends---strictly speaking---on the laser intensity, the dependence is found to be quite weak).
Our analysis suggests that the numerical error that arises from the discretization of the spatial coordinates and the time is notably smaller than the error that arises due to the finite number of channels included in the wave packet decomposition. The impact of the finite number of channels on selected observables  is discussed in the next section.

\subsection{Number of channels}
\label{sec_numerics_convergence}

To analyze the wave packet composition,
we define the, in general, time-dependent probability 
$P^{(J)}_{M}(t)$
to be in the ``rotational channel'' labeled by $J$ and $M$,
\begin{widetext}
\begin{align}
\label{eq_direct}
P^{(J)}_{M}(t)= 
  \int\left|\int \Psi(\rho,\theta,\phi,\beta,\gamma, t)
  \sqrt{\frac{2J+1}{8 \pi^2 }}
  [D^{(J)}_{0,M}(0,\beta,\gamma)]^*dV_{\text{euler}}\right|^2dV_{\text{hyper}}.
\end{align}
\end{widetext}
The time-independent probability ${\cal{P}}^{(J)}$ to be in the
$J$th channel is given by
\begin{eqnarray}
  \label{equation_j_channel}
  {\cal{P}}^{(J)} = \sum_{M=-J,-J+2,\cdots,J} P^{(J)}_M(t).
  \end{eqnarray}
Alternatively, the probability ${\cal{P}}^{(J)}$ can be calculated from the hyperradial weights $F^{(J)}_{m,n}(\rho, t)$ via
\begin{eqnarray}
\label{equation_j_channel_2}
{\cal{P}}^{(J)} = \sum_{m=0,\pm 6,\cdots}\sum_{n=0,1,\cdots} \int |F^{(J)}_{m,n}(\rho,t)|^2 \rho^5 d\rho. 
\end{eqnarray}
Equations~\eqref{equation_j_channel} and \eqref{equation_j_channel_2} provide two ways of calculating ${\cal{P}}^{(J)}$. A comparison of the $t=0^+$ results calculated through direct numerical integration of $\Psi(\rho,\theta,\phi,\beta,\gamma,0^+)$ [Eqs.~\eqref{eq_direct} and \eqref{equation_j_channel}]  and the decomposed $t=0^+$ wave packet [Eq.~\eqref{equation_j_channel_2}] provides an  estimate of the truncation error that results from cutting the infinite double sum over $m$ and $n$ in Eq.~\eqref{equation_j_channel_2} off at $m_{\text{max}}$ and $n_{\text{max}}$, respectively. The dependence of ${\cal{P}}^{(J)}$ on  $m_{\text{max}}$ and $n_{\text{max}}$ is quantified below. In addition,
the variation of the right hand sides of
Eqs.~(\ref{equation_j_channel}) and (\ref{equation_j_channel_2}) with time can be interpreted as an indicator of the accuracy
of our propagation scheme since the right hand side should, theoretically, not vary with
time. 
Indeed, our time propagation scheme is accompanied by  changes of ${\cal{P}}^{(J)}(t)$ below $10^{-5}$ for $t \le 50$~ps. This last statement is fully consistent with our previous conclusion that the main numerical error comes from the channel truncation (see the last paragraph of the previous section). 

\begin{figure}
\includegraphics[width=0.40\textwidth]{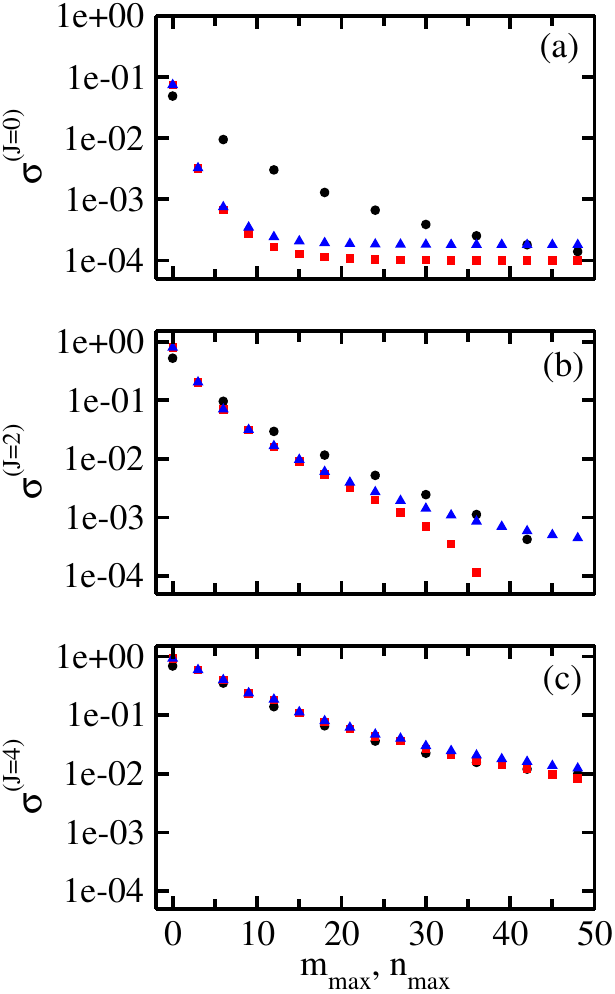}
    \caption{Convergence analysis of $t=0^+$ wave packet decomposition.
    The symbols show the relative error $\sigma^{(J)}$ for
    (a) $J=0$,
    (b) $J=2$, and
    (c) $J=4$.
    Black circles show $\sigma^{(J)}$ as a function of $m_{\text{max}}$ for $m_{\text{min}}=-m_{\text{max}}$ and $n_{\text{max}}=49$.
    Red squares and blue triangles show $\sigma^{(J)}$ as a function of $n_{\text{max}}$ for $-m_{\text{min}}=m_{\text{max}}=60$ and $-m_{\text{min}}=m_{\text{max}}=42$, respectively.
    }
    \label{fig_convergence}
\end{figure}

To  analyze the convergence with respect to the  basis set size, i.e., with respect to $n_{\text{max}}$ and $m_{\text{max}}$, symbols in Figs.~\ref{fig_convergence}(a), \ref{fig_convergence}(b), and \ref{fig_convergence}(c)  
show the relative error $\sigma^{(J)}$,
\begin{eqnarray}
\label{eq_relative_error}
\sigma^{(J)}=
\frac{|{\cal{P}}^{(J)}(\mbox{direct})-{\cal{P}}^{(J)}(\mbox{basis set})|}{{\cal{P}}^{(J)}(\mbox{direct})},
\end{eqnarray}
for $J=0$, $2$, and $4$, respectively. 
In Eq.~(\ref{eq_relative_error}),
${\cal{P}}^{(J)}(\mbox{direct})$ is calculated using Eqs.~(\ref{eq_direct}) and (\ref{equation_j_channel}) and ${\cal{P}}^{(J)}(\mbox{basis set})$ is calculated using Eq.~(\ref{equation_j_channel_2}). 
  To make Fig.~\ref{fig_convergence}, a $\delta$-function
  pulse with $\tau=311$~fs and  $I=3.5 \times 10^{14}$~W/cm$^2$ 
  (corresponding to $\overline{\epsilon}=0.09986$~a.u.) is used~\cite{footnote_kick} and the
  ${\cal{P}}^{(J)}$ are calculated  at $t=0^+$. The
  direct integration  yields essentially exact results within our delta-function model. 
The black circles show the dependence of 
$\sigma^{(J)}$ on $m_{\text{max}}$ 
($m_{\text{min}}=-m_{\text{max}}$) for fixed  $n_{\text{max}}$, namely $n_{\text{max}}=49$, while the red squares and blue triangles show the dependence of 
$\sigma^{(J)}$ on $n_{\text{max}}$ 
 for fixed  $m_{\text{max}}$ and $m_{\text{min}}$, namely $m_{\text{max}}=-m_{\text{min}}=60$ and $m_{\text{max}}=-m_{\text{min}}=42$, respectively. It can be seen that the convergence with increasing $n_{\text{max}}$ and/or $m_{\text{max}}$ is fastest for $J=0$ and slowest for $J=4$. Keeping in mind that the total population is largest for $J=0$ and just $2.5$~\% for $J=4$ (see Table~\ref{table_initial} and the discussion below), a somewhat larger relative error in the  $J=4$ channel compared to the $J=0$ channel should only have a small effect on the overall convergence.
 Figure~\ref{fig_convergence}(a) shows that the population reaches, when $m_{\text{max}}$ is fixed at $60$ or $42$, a plateau at $n_{\text{max}} \approx 20$, with a relative error around a few times $10^{-4}$.
 Since the $J=2$ and $4$ channels show a stronger dependence on $n_{\text{max}}$,
  we chose a somewhat larger value, namely $n_{\text{max}}=29$, for the time-dependent studies. 
 For $m_{\text{max}}$, we choose $42$.

 Table~\ref{table_initial} summarizes the probabilities $P_{M}^{(J)}(0^+)$ and ${\cal{P}}^{(J)}$
  for the basis set used for the time dynamics, namely for 
  $15$ $m$-channels ($-m_{\text{min}}= m_{\text{max}}=42$)
and $30$ $n$-channels ($n_{\text{max}}=29$) for fixed $J$,
resulting in $15\times30\times3$ coupled equations that are being propagated. The factor of $3$ arises since we  account for the $J=0$, $2$, and $4$ wave packet components.
  It can be seen that the dominant partial wave channels are $J=0$ and $2$,
  with the $J=4$ channel contributing $2.5$~\%. Using $J=0-4$, the $0^+$ wave packet captures 99.5~\% of the initial state.
  This number, combined with the differences between the probabilities calculated through direct integration, provides an estimate of the truncation error.

 \begin{table}
    \begin{tabular}{c|r|c| c|c}
        $J$ & $M$  & $P_{M}^{(J)}(0^+)(\mbox{direct})$ & ${\cal{P}}^{(J)}(\mbox{direct})$ & ${\cal{P}}^{(J)}(\mbox{basis set})$ \\ \hline
        $0$ & $0$ & $0.8373$ & $0.8373$ & $0.8372$\\ \hline
        $2$ & $-2$ & $0.03599$ && \\
        $2$ & $0$ & $0.06132$&&\\
        $2$ & $2$ & $0.03599$ & $0.1334$&$0.1333$\\ \hline
        $4$ & $-4$ & $0.00464$&&\\
        $4$ & $-2$ & $0.00410$&&\\
        $4$ & $0$ & $0.00710$&&\\
        $4$ & $2$ & $0.00410$&&\\
                $4$ & $4$ & $0.00464$& $0.0246$& $0.0243$
        \end{tabular}
        \caption{Decomposition of the $t=0^+$ wave packet for a laser with intensity $3.5\times 10^{14}$~W/cm$^2$ and $\tau=331$~fs. The numbers in the  third and fourth columns  are obtained through direct integration; this approach is essentially exact. The numbers in the fifth column are obtained using a basis set with $m_{\text{max}}=-m_{\text{min}}=42$ and $n_{\text{max}}=29$. 
        This basis  set captures $99.5$~\% of the $t=0^+$ wave packet.
        The difference between the probabilities obtained by the two methods provides an estimate of the truncation error. 
     }
        \label{table_initial}
        \end{table}

  Recall from our earlier discussion that the $J \ge 2$ wave packet components are
  unbound. If we were to decompose them into eigen states of
  $H_{\text{trimer,rel}}$, they would be spanned by the unbound field-free
  trimer eigen states (scattering eigen states of  the field-free
  trimer).
  The $J=0$ scattering portion of the wave packet is defined through 
  \begin{eqnarray}
    \label{eq_wave_scatt_j0}
  \Psi^{(0,\text{scatt})}(\rho,\theta,\phi,\beta,\gamma,t)
  = \nonumber \\
  \Psi^{(0)}(\rho,\theta,\phi,\beta,\gamma,t)
  - \nonumber \\
  c_{\text{trimer}}^{(\text{ground})}
  \exp \left(-\imath \frac{E_{\text{trimer}}^{(\text{ground})} t }{\hbar} \right)
  \sqrt{\frac{1}{8 \pi^2}}
  \psi_{\text{trimer}}^{(\text{ground})}(\rho,\theta,\phi)- \nonumber \\
    c_{\text{trimer}}^{(\text{Efimov})}
  \exp \left(-\imath \frac{E_{\text{trimer}}^{(\text{Efimov})} t }{\hbar} \right)
  \sqrt{\frac{1}{8 \pi^2}}
  \psi_{\text{trimer}}^{(\text{Efimov})}(\rho,\theta,\phi) 
  .\nonumber \\
\end{eqnarray}
  It is instructive to project 
  $\Psi(\rho,\theta,\phi,\beta,\gamma,t)$
  onto the bound eigen states $\psi_{\text{trimer}}^{(\text{ground})}(\rho,\theta,\phi)$
  and $\psi_{\text{trimer}}^{(\text{Efimov})}(\rho,\theta,\phi)$;
  the latter is the Efimov trimer (bound first excited state) with energy
  $E_{\text{trimer}}^{(\text{Efimov})}$.
  We find that, respectively, $80.4$~\% ($|c_{\text{trimer}}^{(\text{ground})}|^2=0.804$) and $0.00412$~\% ($|c_{\text{trimer}}^{(\text{Efimov})}|^2=0.0000421$) of the $\Psi(\rho,\theta,\phi,\beta,\gamma,0^+)$ component are in the ground eigen state and first excited eigen state.
  This implies that the $J=0$ component has a probability of $3.32$~\% 
  ($|c_{\text{trimer}}^{(\text{unbound})}|^2=0.0332$) 
  to
  be in a scattering state (this probability is time-independent). Table~\ref{table_initial2} compares these values, obtained from direct integration, with those obtained from the basis set used to perform the time propagation. 
  For comparison, the same laser pulse for the dimer promotes $1.39$~\%, $1.04$~\%, and $0.124$~\%
  into the unbound $J=0$, $2$, and $4$ components, respectively.
  The dependence of the ground state population 
  $|c_{\text{trimer}}^{(\text{ground})}|^2$ on the laser intensity is discussed in Sec.~\ref{sec_results}.

\begin{table}
    \begin{tabular}{c|c|c}
      &  direct  & basis set \\
     \hline
     $|c_{\text{trimer}}^{\text{(ground)}}|^2$ & 0.8041 & 0.8039\\
     $|c_{\text{trimer}}^{\text{(Efimov)}}|^2$ & $0.4121\times10^{-4}$ & $0.4115\times10^{-4}$ \\
     $|c_{\text{trimer}}^{\text{(unbound)}}|^2$ & $0.3322\times10^{-1}$ & $0.3333\times10^{-1}$ \\
        \end{tabular}
       \caption{Decomposition of the $J=0$ component of the $t=0^+$ wave packet for a laser with intensity $3.5\times 10^{14}$~W/cm$^2$ and $\tau=331$~fs. The difference between the probabilities obtained by the two methods provides an estimate of the truncation error. 
        }
        \label{table_initial2}
        \end{table}

\section{Observables}
\label{sec_observables}

The time-dependent expectation value $\langle A\rangle(t)$ of the operator
$A$ with respect to the time-dependent
wave packet $\Psi(\rho,\theta,\phi,\beta,\gamma,t)$
is given by
\begin{widetext}
\begin{eqnarray}
  \langle A \rangle (t) =
  \int [\Psi(\rho,\theta,\phi,\beta,\gamma,t)]^*
A
\Psi(\rho,\theta,\phi,\beta,\gamma,t) dV_{\text{hyper}} dV_{\text{euler}},
\end{eqnarray}
where it is
 assumed that the wave packet $\Psi(\rho,\theta,\phi,\beta,\gamma,t)$ is normalized to $1$ for all $t$.
In addition to  the expectation value with respect to the full wave
packet, we also define the expectation value
$\langle A \rangle^{(J)} (t)$ with respect to the $J$th
partial wave component,
\begin{eqnarray}
  \langle A \rangle^{(J)} (t) =
  \frac{
    \int [\Psi^{(J)}(\rho,\theta,\phi,\beta,\gamma,t)]^*
A
\Psi^{(J)}(\rho,\theta,\phi,\beta,\gamma,t) dV_{\text{hyper}} dV_{\text{euler}}
  }{
    \int |\Psi^{(J)}(\rho,\theta,\phi,\beta,\gamma,t)|^2
    dV_{\text{hyper}} dV_{\text{euler}}},
\end{eqnarray}
and with respect to 
 the scattering portion
$\Psi^{(0,\text{scatt})}(\rho,\theta,\phi,\beta,\gamma,t)$
 of the $J=0$ channel,
 \begin{eqnarray}
  \langle A \rangle^{(0,\text{scatt})} (t) =
  \frac{
    \int [\Psi^{(0,\text{scatt})}(\rho,\theta,\phi,\beta,\gamma,t)]^*
A
\Psi^{(0,\text{scatt})}(\rho,\theta,\phi,\beta,\gamma,t) dV_{\text{hyper}} dV_{\text{euler}}
  }{
    \int |\Psi^{(0,\text{scatt})}(\rho,\theta,\phi,\beta,\gamma,t)|^2
    dV_{\text{hyper}} dV_{\text{euler}}}.
\end{eqnarray}
\end{widetext}
  
Expectation values of interest include the average internuclear distance
$r_{\text{ave}}=\langle A_{\text{pair,ave}} \rangle(t)$, where
\begin{eqnarray}
  A_{\text{pair,ave}} = \frac{1}{3} \left(r_{12}+r_{13}+r_{23} \right),
\end{eqnarray}
the average of the smallest  internuclear	distance
$r_{\text{min}}=\langle A_{\text{pair,min}} \rangle(t)$, where
\begin{eqnarray}
  A_{\text{pair,min}} = \text{min}\left(r_{12},r_{13},r_{23} \right),
\end{eqnarray}
and the average of the largest  internuclear	distance
$r_{\text{max}}=\langle A_{\text{pair,max}} \rangle(t)$, where
\begin{eqnarray}
  A_{\text{pair,max}} = \text{max}\left(r_{12},r_{13},r_{23} \right).
\end{eqnarray}
In addition to these ``pair correlators,'' we monitor the ``trimer geometry'' at fixed time $t$ using a coordinate system defined by the principal moments of inertia~\cite{Nielsen_1998}.

Another important observable is the
expectation value $\text{KER}(t)=\langle A_{\text{KER}}\rangle(t)$ of
the kinetic energy release operator
$A_{\text{KER}}$~\cite{doi:10.1126/science.aaa5601},
\begin{eqnarray}
A_{\text{KER}} =   \frac{e^2}{4 \pi \epsilon_0} 
\left( \frac{1}{r_{12}}+\frac{1}{r_{13}}+\frac{1}{r_{23}}
\right).
\end{eqnarray}
   It describes the trimer after the application of a probe laser pulse, which ionizes each of the atoms. Note that the energy scale of the KER after the pump pulse is on the order of several eV, at least four orders of magnitude larger than the kinetic energy of the kicked neutral system. 
   It is this separation of scales that makes the spatial  coordinates discussed in this work accessible to experiment using Coulomb explosion imaging.
    The probability distribution $P({\text{KER}},t)$ of $\text{KER}(t)$
    is obtained through
    \begin{eqnarray}
      \label{eq_histo_ker}
  P(\text{KER}, t) = \nonumber \\
    \int |\Psi(\rho,\theta,\phi,\beta,\gamma,t)|^2
\delta \left(\text{KER}-  A_{\text{KER}}
  \right) 
  dV_{\text{hyper}} dV_{\text{euler}}. \nonumber \\
      \end{eqnarray}
   The quantities   $P({\text{KER}},t)$ and ${\text{KER}}(t)$ can be measured experimentally
   without the need of processing experimental data with the aid of a reconstruction algorithm~\cite{doi:10.1126/science.aaa5601}.
In terms of the partial wave decomposition
[see Eqs.~(\ref{eq_wave_j}) and (\ref{eq_wave_scatt_j0})],
we obtain 
\begin{widetext}
    \begin{eqnarray}
      \label{eq_histo_ker_partial}
         P(\text{KER}, t)
         =
         |c_{\text{trimer}}^{(\text{ground})}|^2
         P_{\text{trimer}}^{(\text{ground})}(\text{KER}) +
         P^{(0,\text{scatt})}(\text{KER},t)
         +
         \sum_{J=2,4,\cdots} P^{(J)}(\text{KER},t)
         + \nonumber \\
\left( c_{\text{trimer}}^{(\text{ground})} \right)^* 
\exp \left(\imath \frac{E_{\text{trimer}}^{(\text{ground})} t }{\hbar} \right)
\int [\psi_{\text{trimer}}^{(\text{ground})}(\rho,\theta,\phi)]^*
\overline{\Psi}^{(0,\text{scatt})}(\rho,\theta,\phi,t)
\delta \left(\text{KER}-   A_{\text{KER}}
  \right) 
dV_{\text{hyper}} + \mbox{c.c.}
      \end{eqnarray}
      \end{widetext}
      for the helium trimer. 
    In writing Eq.~(\ref{eq_histo_ker_partial}), we
    suppressed terms that contain $\psi_{\text{trimer}}^{(\text{Efimov})}(\rho,\theta,\phi)$
    due to this state's neglegibly small occupation
    and defined $\overline{\Psi}^{(0,\text{scatt})}(\rho,\theta,\phi,t)=\sqrt{8 \pi^2} \Psi^{(0,\text{scatt})}(\rho,\theta,\phi,\beta,\gamma,t)$.
    Note that $P_{\text{trimer}}^{(\text{ground})}(\text{KER})$
    is independent of time,
    \begin{eqnarray}
      \label{eq_histo_ker_ground}
  P^{(\text{ground})}_{\text{trimer}}(\text{KER}) =\nonumber \\
  \int |\psi_{\text{trimer}}^{(\text{ground})}(\rho,\theta,\phi)|^2
\delta \left(\text{KER}-   
A_{\text{KER}}
\right)
  dV_{\text{hyper}}.
    \end{eqnarray}
    For the helium trimer, $P_{\text{trimer}}^{(\text{ground})}(\text{KER})$
    has
    been
        determined both experimentally and theoretically in earlier work~\cite{doi:10.1126/science.aaa5601}.
          In Eq.~(\ref{eq_histo_ker_partial}), the probability distributions
          $P^{(J)}({\text{KER}},t)$ and $P^{(0,\text{scatt})}({\text{KER}},t)$ are defined
          by replacing $\Psi(\rho,\theta,\phi,\beta,\gamma,t)$ in Eq.~(\ref{eq_histo_ker})
          by $\Psi^{(J)}(\rho,\theta,\phi,\beta,\gamma,t)$ and
          $\Psi^{(0,\text{scatt})}(\rho,\theta,\phi,\beta,\gamma,t)$, respectively, and correspondingly normalizing the distribution function with respect to which the probability function is being calculated.

 The hyperradius-resolved 
 expectation value $\langle A \rangle(\rho,t)$,
i.e., the $\rho$- and $t$-resolved expectation value of $A$, is defined through
\begin{widetext}
\begin{eqnarray}
  \langle A \rangle(\rho,t)
  =\frac{\int [\Psi(\rho,\theta,\phi,\beta,\gamma,t)]^*
    A
\Psi(\rho,\theta,\phi,\beta,\gamma,t) 
\sin(4 \theta) d \theta d \phi
dV_{\text{euler}}}{\int |\Psi(\rho,\theta,\phi,\beta,\gamma,t)|^2
 \sin(4 \theta) d \theta d \phi
dV_{\text{euler}}}.
\end{eqnarray}
\end{widetext}
To characterize the ``orientational dynamics,''
we calculate the alignment signal $\langle \cos ^2 (\beta) \rangle(\rho,t)$~\cite{schouder2022}. 
Since $\beta$ is the angle between the polarization axis and
the vector that is normal to the trimer plane, we have the following:
For $\beta=0$, the trimer plane is perpendicular to the polarization axis.
For $\beta=\pi/2$, in turn, the trimer plane is parallel to the polarization axis.
Since an isotropic wave packet leads to 
$\langle \cos ^2 (\beta) \rangle(\rho,t)=1/3$, i.e., a $\rho$- and $t$-independent value,
a deviation from $1/3$ (and associated $\rho$- and $t$-dependence)
provides a fingerprint of the laser-trimer interaction.
This can be seen explicitly by rewriting $\cos^2 (\beta)$ in terms of the Wigner-D
functions,
\begin{eqnarray}
  \cos^2 (\beta) =
  \frac{1}{3} D_{0,0}^{(0)}(\alpha,\beta,\gamma)+\frac{2}{3} D_{0,0}^{(2)}
  (\alpha,\beta,\gamma).
  \end{eqnarray}
Using $D_{0,0}^{(0)}(\alpha,\beta,\gamma)=1$ and orthogonality of the Wigner-D functions,
it can be seen readily that an isotropic $J=0$ wave packet
yields $\langle \cos ^2 (\beta) \rangle(\rho,t)=1/3$.
If, in contrast, the wave packet contains $J\ge 2$ components, then $\langle \cos ^2 (\beta) \rangle(\rho,t)$
deviates from $1/3$ and displays, in general, a dependence on the hyperradius
$\rho$ and the time $t$.
It should be noted that the trimer alignment signal has similarities with the
alignment signal
for the helium dimer~\cite{nature-physics}, with $\rho$ replaced by the internuclear distance and
$\beta$ replaced by the 
angle between the polarization axis and the molecular axis of the dimer.

To higlight that the alignment signal of the trimer can, analogously to
that of the dimer~\cite{nature-physics},
be interpreted as
an interference between different $J$ wave packet components,
we evaluate the alignment signal,
using the partial wave decomposition given in Eq.~(\ref{eq_expand}).
This yields
\begin{widetext}
\begin{eqnarray}
  \label{eq_interference}
  \langle \cos ^2 (\beta) \rangle(\rho,t) =
  \frac{1}{3} +
\frac{2}{3\sum_{J=0,2,\cdots}\sum_{m=0,\pm 6,\cdots}\sum_{n=0,1,\cdots}\left|F^{(J)}_{m,n}(\rho, t)\right|^2}\times\\ \nonumber
\sum_{\substack{J,J'=0,2,\cdots \\ |J'-J|\le 2}} \sum_{m=0, \pm 6,\cdots}
   \sum_{ n',n=0,1,\cdots} \mathcal{G}^{(J',J)}_{m,n',n}|F^{(J')}_{m,n'}(\rho,t)
     F^{(J)}_{m,n}(\rho,t)|
     \cos \left[ \delta^{(J')}_{m,n'}(\rho,t) - \delta^{(J)}_{m,n}(\rho,t)\right]
     , 
\end{eqnarray}
\end{widetext}
where the phase $\delta^{(J)}_{m,n}(\rho,t)$ arises from
writing the weight functions $F^{(J)}_{m,n}(\rho,t)$ in terms of their magnitude and
phase factor,
\begin{eqnarray}
  F^{(J)}_{m,n}(\rho,t) = |F^{(J)}_{m,n}(\rho,t)| \exp \left[ \imath
  \delta^{(J)}_{m,n}(\rho,t) \right].
\end{eqnarray}
The constant $1/3$ 
on the right hand side of Eq.~(\ref{eq_interference}) arises from the
  $D_{0,0}^{(0)}$ contribution to $\cos^2(\beta)$
while the second term (sums over $J'$, $J$, $m$,  $n'$,  and $n$)
on the right hand side of Eq.~(\ref{eq_interference}) arises from	the
  $D_{0,0}^{(2)}$ contribution to $\cos^2(\beta)$.
The coefficients 
${\cal{G}}^{(J',J)}_{m,n',n}$
arise from the integration over the  angles, 
 \begin{widetext}
\begin{align}
\label{G_integral}
  {\cal{G}}^{(J',J)}_{m,n',n} &
  = 
  \int \left[\Phi^{(J')}_{m,n'}(\theta,\phi,\beta,\gamma)\right]^*
  D_{0,0}^{(2)}(\alpha,\beta,\gamma)\Phi^{(J)}_{m,n}(\theta,\phi,\beta,\gamma)\frac{1}{4}\sin(4\theta)d\theta d\phi dV_{\text{euler}} .
\end{align}
Defining
\begin{eqnarray}
{\cal{D}}_{M}^{(J',J)}=\sqrt{\frac{(2J+1)(2J'+1)}{16\pi^2}}\int [D_{0,M}^{(J')}(0,\beta,\gamma)]^*
   D_{0,0}^{(2)}(0,\beta,\gamma)
    D_{0,M}^{(J)}(0,\beta,\gamma)
    \sin(\beta) d \beta d \gamma,
\end{eqnarray}
one finds
\begin{align}
 {\cal{G}}^{(J',J)}_{m,n',n} 
  = 
  \sum_{M=-\text{min}(J,J')}^{\text{min}(J,J')}\int [P_{M,n'}^{(J',m)}(\theta)]^* 
  P_{M,n}^{(J,m)}(\theta){\cal{D}}^{(J',J)}_{M}
  \frac{1}{4}\sin(4 \theta) d \theta,
  \end{align}
\end{widetext}
where the notation ``$\text{min}(J,J')$'' means the minimum value of $J$ and $J'$.
In writing Eq.~(\ref{eq_interference}), we assumed that the
$P_{M,n}^{(J,m)}(\theta)$ are real.
This can be accomplished by chosing the phase factors of $P_{M,n}^{(J,m)}(\theta)$
accordingly.
Keeping in mind that
the coefficients ${\cal{D}}^{(J',J)}_{M}$ and ${\cal{G}}^{(J',J)}_{m,n',n}$ are real,
the second term on the right hand side of Eq.~(\ref{eq_interference})
can be identified as an interference term, with the phase difference at each $\rho$ being set by the arguments of the weight functions; an interpretation within a minimal basis is presented in Sec.~\ref{sec_results}.

Last, to make direct contact with the alignment dynamics of the helium dimer, we calculate the distance- and time-resolved expectation values $\langle \cos^2(\vartheta_{\text{ave}}) \rangle(r,t) $, 
$\langle \cos^2(\vartheta_{\text{min}}) \rangle(r,t) $, 
and
$\langle \cos^2(\vartheta_{\text{max}}) \rangle(r,t) $,
where
\begin{widetext}
        \begin{eqnarray}
          \label{eq_dimer_connect1}
          \langle\cos^2(\vartheta_{\text{ave}}) \rangle(r,t)  =\frac{\frac{1}{3} \sum_{j=1}^2 \sum_{k=j+1}^3
            \int |\Psi(\rho,\theta,\phi,\beta,\gamma,t)|^2 \cos^2(\vartheta_{jk})
            \frac{\delta(r_{jk}-r)}{r^2}
            dV_{\text{hyper}} dV_{\text{euler}}}{\frac{1}{3}\sum_{j=1}^2 \sum_{k=j+1}^3
            \int |\Psi(\rho,\theta,\phi,\beta,\gamma,t)|^2
            \frac{\delta(r_{jk}-r)}{r^2}dV_{\text{hyper}} dV_{\text{euler}}}
               \end{eqnarray}
               and
                 \begin{eqnarray}
                  \label{eq_dimer_connect2}
          \langle\cos^2(\vartheta_{\text{min}}) \rangle(r,t)  =\frac{
            \int |\Psi(\rho,\theta,\phi,\beta,\gamma,t)|^2 \cos^2(\vartheta_{\text{min}})
            \frac{\delta(r_{\text{min}}-r)}{r^2}
            dV_{\text{hyper}} dV_{\text{euler}}}{\int |\Psi(\rho,\theta,\phi,\beta,\gamma,t)|^2
            \frac{\delta(r_{\text{min}}-r)}{r^2}
            dV_{\text{hyper}} dV_{\text{euler}}}.          
               \end{eqnarray}
               \end{widetext}
        In Eq.~(\ref{eq_dimer_connect1}),
the angle $\vartheta_{jk}$ denotes, just as earlier, the angle between the internuclear distance vector $\vec{r}_{jk}$ and the space-fixed $z$-axis (direction of the linear laser polarization).   In Eq.~(\ref{eq_dimer_connect2}),   
$\vartheta_{\text{min}}$ is the angle between the vector $\vec{r}_{\text{min}}$, defined through $r_{\text{min}}=\text{min}(r_{12},r_{13},r_{23})$, and the space-fixed $z$-axis.
The quantity $\langle \cos ^2 (\vartheta_{\text{max}} )\rangle (r,t)$ is defined analogously to $\langle \cos ^2 (\vartheta_{\text{min}} )\rangle (r,t)$, 
with the subscript ``$\text{min}$" replaced by the subscript ``$\text{max}$."

Since some operators such as, e.g., $A_{\text{pair,max}}$, do not have compact analytical expressions, we resort to determining 
expectation values through Monte Carlo sampling~\cite{doi:10.1063/1.1699114}.
For the determination of $\langle A \rangle(t)$,
$\langle A \rangle(\rho,t)$, and $\langle A \rangle(r,t)$, the Monte Carlo walk at fixed time is guided by the
probability distribution
\begin{eqnarray}
\frac{|\Psi(\rho,\theta,\phi,\beta,\gamma,t)|^2}
{\int |\Psi(\rho,\theta,\phi,\beta,\gamma,t)|^2 dV_{\text{hyper}} dV_{\text{euler}}}.
\end{eqnarray}
Configurations are generated by sampling the configuration space using the 
hyperspherical coordinates $\rho$, $\theta$, and $\phi$ and the Euler angles $\beta$ and $\gamma$.
For observables whose expectation value is calculated with respect to a specific $J$ component or the scattering portion of the $J=0$ component, the sampling is changed accordingly.

\section{Helium trimer dynamics}
\label{sec_results}

This section reports on the dynamics of the bosonic helium trimer. As mentioned in the previous sections, the static properties of the helium trimer have attracted significant attention since the two $J=0$ trimer bound states are diffuse.
Throughout this section we use, as in the convergence analysis presented in Sec.~\ref{sec_numerics}, a pulse with a duration of $\tau=311$~fs  and intensity $I=3.5 \times 10^{14}$~W/cm$^2$. The initial state is given by 
Eq.~(\ref{wave_initial}).

\begin{figure}
    \includegraphics[width=0.4\textwidth]{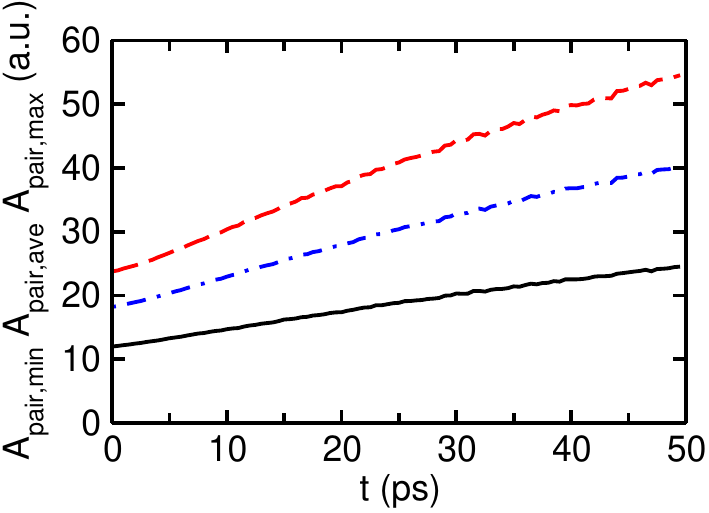}
    \caption{Dynamics of internuclear distances. The black solid, blue dash-dotted, and red dashed  lines show   $A_{\text{pair,min}}$, 
    $A_{\text{pair,ave}}$, and $A_{\text{pair,max}}$, 
    respectively, as a function of time.
     }
    \label{fig_result_pair}
\end{figure}

The blue dash-dotted line in Fig.~\ref{fig_result_pair} shows the time evolution of the expectation value $A_{\text{pair,ave}}$ of the pair distance. The pair distance increases monotonically with increasing time. This behavior is consistent with the occupation probabilities at $t=0^+$. The portion of the wave packet that occupies the ground eigen state is characterized by a constant pair distance while the population of the scattering states with $J=0$ and $J>0$  are characterized by  pair distances that increase with increasing time. Figure~\ref{fig_result_pair} shows that $A_{\text{pair,ave}}$ has roughly doubled after $50$~ps (this is the largest time considered in Fig.~\ref{fig_result_pair}). To place this time scale into context, we note that the depth of the helium-dimer potential ($\approx 11$~K) corresponds to 
$4.36$~ps 
while the trimer ground state energy ($131$~mK) corresponds to $366$~ps~\cite{footnote_conversion}. For comparison, the black solid and red dashed lines in Fig.~\ref{fig_result_pair} show the time evolution of the expectation values of the smallest and largest interparticle distances. Both increase with time.

\begin{figure}
    \includegraphics[width=0.5\textwidth]{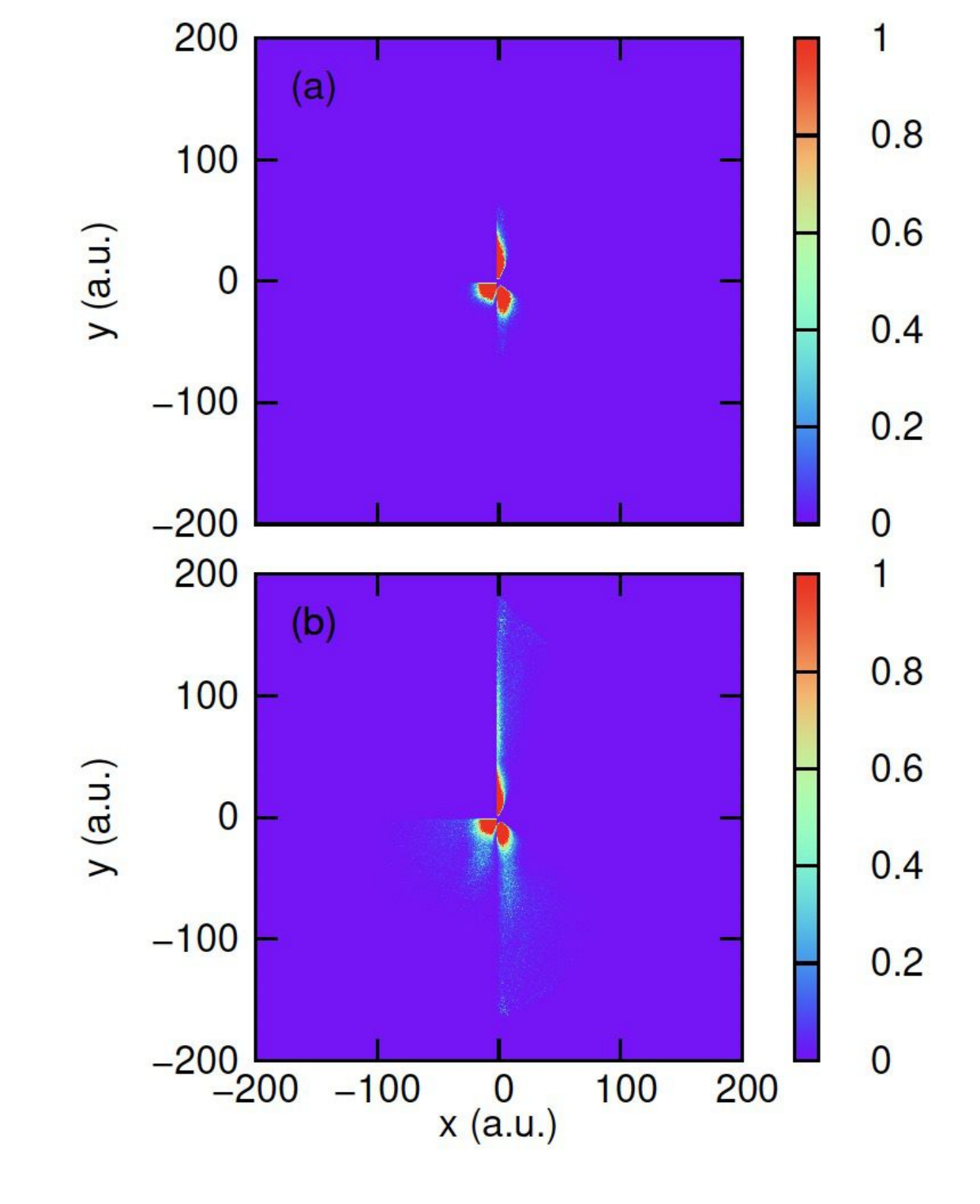}
    \vspace{-0.2cm}
    \caption{Snapshot at (a) $5$~ps and (b) $40$~ps of the trimer geometries using the moment of inertia axes frame representation introduced by Nielsen {\em{et al.}}~\cite{Nielsen_1998}.  The legends on the right color code the density (i.e., the number of counts); an arbitrary normalization is used in both panels. To highlight the lower count regions, the peak value is saturated at $1$. The supplemental material contains a movie covering the time interval  $0 \le t \le 50$~ps.  }
\label{fig_result_nielsen}
\end{figure}

To obtain a detailed understanding of how the trimer geometries change with time, Fig.~\ref{fig_result_nielsen} shows snapshots at two different times (early during the time evolution, i.e., at $5$~ps, and later during the time evolution, i.e., at $40$~ps) using the trimer orientation in the moments-of-inertia coordinate system  introduced by Nielsen {\em{et al.}}~\cite{Nielsen_1998}. The snapshot in Fig.~\ref{fig_result_nielsen}(b) shows that the trimer wave packet contains configurations where all three particles are comparatively far apart from each other. The supplemental material~\cite{SM} contains a movie of the trimer evolution (time resolution of $0.5$~ps for $t \le 50$~ps), using the same
representation.
The movie provides visual evidence that the early time dynamics is characterized by appreciable rearrangement of configurations while the later time dynamics is characterized predominantly by self-similar expansion. This aspect of the dynamics 
will be  
characterized in detail in a separate publication~\cite{our-experiment-theory-paper}.

\begin{figure}
    \includegraphics[width=0.4\textwidth]{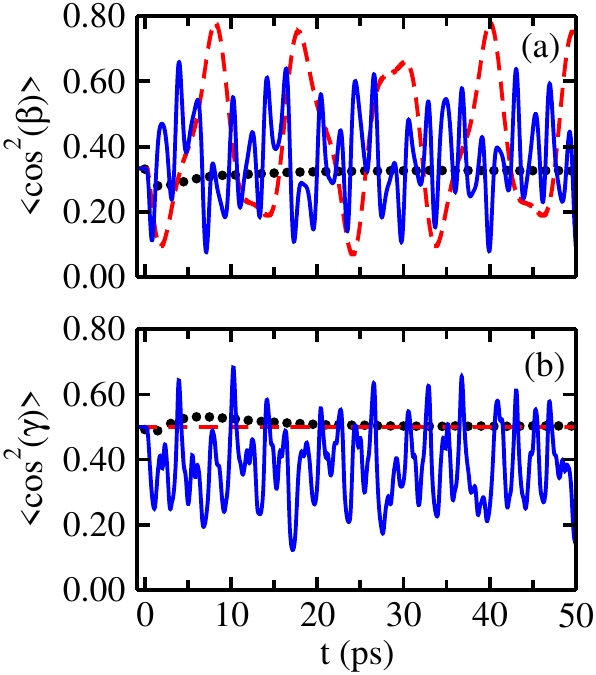}
    \caption{Averaged   orientational dynamics. The black circles show (a) $\langle \cos^2(\beta)\rangle(t)$ and (b) 
$\langle \cos^2(\gamma)\rangle(t)$ as a function of time. These orientation measures are equal to, respectively, $1/3$ and $1/2$ for an isotropic trimer. For comparison, the red dashed and blue solid lines show results from the rigid-body framework introduced in Appendix~\ref{appendix_rigid}  
for configurations characterized by  $r_{12}=7.833$~a.u. and $r_{13}=r_{23}=7.478$~a.u. and by $r_{12}=r_{13}=8.643$~a.u. and $r_{23}=4.879$~a.u., respectively.
  }    \label{fig_result_average_euler}
\end{figure}

To investigate in which way, if any, the laser pulse leads to a preferred orientation or alignment of the trimer, we analyze
expectation values of the Euler angles, namely the expectation values  $\langle \cos^2(\beta)\rangle(t)$ and
$\langle \cos^2(\gamma)\rangle(t)$.
For an isotropic wave packet, these expectation values are equal to $1/3$ and $1/2$.
The black circles in Figs.~\ref{fig_result_average_euler}(a) and \ref{fig_result_average_euler}(b) show that
 $\langle \cos^2(\beta)\rangle(t)$ and
$\langle \cos^2(\gamma)\rangle(t)$ deviate
by up to 15~\% from $1/3$ and $1/2$, respectively, for $t \lesssim 20$~ps. For larger times, both quantities are essentially constant, which is consistent with an, on average,  isotropic helium trimer distribution.
The observed behavior is very different from what one  observes for a rigid body.
Using the framework introduced in Appendix~\ref{appendix_rigid}, the red dashed and blue solid   lines in Figs.~\ref{fig_result_average_euler}(a) and \ref{fig_result_average_euler}(b) show the results for rigid bodies whose shapes are fairly close to an equilateral triangle and a linear chain, respectively.
The former displays approximately five distinct oscillations over the 50~ps time interval shown while the latter displays more than 30. For the nearly equilateral triangle, the hyperangle $\theta$ is close to $\pi/4$, implying that the $\tan^2(2\theta)J^2_{\text{op},z}$ term in Eq.~(\ref{eq_kinetic}) dominates. 
In this case, the coupled rigid-body equations converge with just a few $J$ channels
 and the oscillation period of the red dashed line in Fig.~\ref{fig_result_average_euler} is, to a good approximation, governed by the difference between the two energetically lowest-lying rigid-body eigen energies (the energy difference corresponds to a time scale of $10.6$~ps). 
For the nearly linear trimer, in contrast, the hyperangle $\theta$ is close to $0$, implying that the $\cos(2\theta)(J^2_{\text{op},+}+J^2_{\text{op},-})$ term in Eq.~(\ref{eq_kinetic}) dominates. In addition, the $\sin^2(2\theta)$ term in the denominator in Eq.~(\ref{eq_kinetic}) is very large, leading to a vastly reduced time scale. 
In this case, convergence of the coupled rigid-body   equations requires a comparatively large number of $J$ channels.
From the rigid-body dynamics shown in Fig.~\ref{fig_result_average_euler}, it might be speculated that a weighted average over different rigid-body configurations could explain the essentially featureless behavior of the orientational measures $\langle \cos^2 (\beta) \rangle(t)$ and  
$\langle \cos^2 (\gamma) \rangle(t)$.

\begin{figure}
\includegraphics[width=0.45\textwidth]{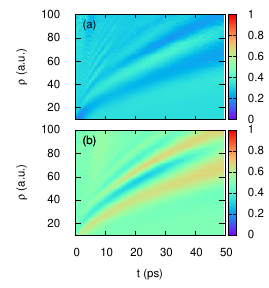} 
 \vspace{-0.2cm}
    \caption{Hyperradius-resolved dynamics of the trimer orientation (``rotational dynamics'').
     Alignment (a) $\langle \cos^2(\beta)\rangle (\rho,t)$ and (b) $\langle \cos^2( \gamma)\rangle (\rho,t)$ as functions of the hyperradius $\rho$ and the time $t$. The  legends on the right color code  the expectation values. For reference, the expectation values $\langle \cos^2(\beta)\rangle (\rho,t)$ and  $\langle \cos^2( \gamma)\rangle (\rho,t)$ for the isotropic function $D_{0,0,}^{(0)}(\alpha,\beta,\gamma)$ are $1/3$ and $1/2$, respectively.  }
\label{fig_result_dynamics_external}
\end{figure}

\begin{figure}
\includegraphics[width=0.45\textwidth]{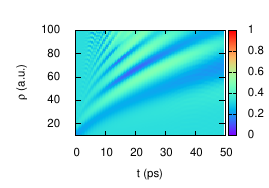} 
 \vspace{-0.2cm}
    \caption{Alignment signal $\langle\cos^2(\beta)\rangle(\rho, t)$ calculated using the $(J=0, m=0, n=0)$ and $(J=2, m=0, n=0)$ channels. 
    The  legend on the right color codes  the alignment signal $\langle\cos^2(\beta)\rangle(\rho, t)$.
    While the time propagation utilizes many 
    channels, the expectation value is calculated using just two channels.
}
\label{two_state_model}
\end{figure}

Motivated by the preceding discussion and  by the fact that the alignment signal of the helium dimer displays distinct signatures of partial wave interference~\cite{nature-physics}, Figs.~\ref{fig_result_dynamics_external}(a)
and \ref{fig_result_dynamics_external}(b)
show the hyperradius-resolved expectation values
$\langle \cos^2(\beta)\rangle (\rho,t)$ and 
     $\langle \cos^2( \gamma)\rangle (\rho,t)$, respectively. Both quantities show  finger-like patterns, i.e., regions with values lower and higher than the average that move to larger $\rho$ with increasing $t$. The quantity $\langle \cos^2(\beta)\rangle (\rho,t)$, e.g., alternates at fixed time between regions with values larger than $1/3$ and with values lower than $1/3$ as $\rho$ increases. This oscillatory behavior exists over the entire time interval considered, i.e., for $t \lesssim 20$~ps, where the average $\langle \cos^2 (\beta) \rangle(t)$ deviates from $1/3$, and for $t \gtrsim 20$~ps, where  $\langle \cos ^2(\beta) \rangle(t)$ is essentially equal to $1/3$. The finger-like patterns are the result of an interference of the $J=0$ and $J=2$ wave packet components. We arrive at this conclusion since the full dynamics
of $\langle \cos ^2 (\beta) \rangle (\rho,t)$ [see Fig.~\ref{fig_result_dynamics_external}(a)] is reproduced semi-quantitatively if the expectation values are calculated using just two basis states, namely 
       the 
$(J,m,n)=(0,0,0)$ and 
$(2,0,0)$
channels; the result is shown in Fig.~\ref{two_state_model}.

Using the reduced basis, the alignment signal reads
\begin{eqnarray}
  \label{eq_interference_reduced}
  \langle \cos ^2 (\beta) \rangle(\rho,t) = 
  \frac{1}{3}
     +
      \frac{4}{3}\frac{
    {\cal{G}}^{(0,2)}_{0,0,0}    
    |F^{(0)}_{0,0}(\rho,t)F^{(2)}_{0,0}(\rho,t)|}{|F^{(0)}_{0,0}(\rho,t)|^2+|F^{(2)}_{0,0}(\rho,t)|^2}\times  \nonumber \\ \cos \left[ \delta^{(2)}_{0,0}(\rho,t) - \delta^{(0)}_{0,0}(\rho,t) \right].
\end{eqnarray}
Analyzing the different terms, it is found that the finger-like pattern can be attributed to the $\rho$- and $t$-dependent phase difference $\delta^{(2)}_{0,0}(\rho,t) - \delta^{(0)}_{0,0}(\rho,t)$ between the $J=0$ and $J=2$ partial wave components, i.e., the fingerlike pattern is a signature of an interference between two different partial wave components. 
It is worthwhile pointing out that the approximate wave packet used in this analysis is independent of $\phi$ since  the reduced basis is characterized by $m=0$.

\begin{figure}
\vspace*{-0.3cm}
\includegraphics[width=0.45\textwidth]{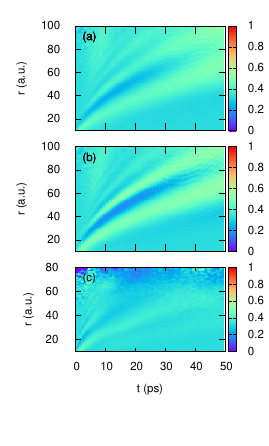}
 \vspace{-1.3cm}
    \caption{Internuclear distance-resolved dynamics of the trimer orientation (this quantity facilitates comparison with the dimer dynamics).
    Expectation values (a) $\langle \cos^2(\vartheta_{\text{ave}})\rangle (r,t)$, 
    (b) $\langle \cos^2(\vartheta_{\text{max}})\rangle (r,t)$, and
    (c) $\langle \cos^2(\vartheta_{\text{min}})\rangle (r,t)$ as functions of the distance $r$ and the time $t$. The  legends on the right color code  the expectation values. Note that the $y$-axes in (a) and (b) extends to $r=100$~a.u. while that in (c) extends only to $r=80$~a.u. [for (c), a smaller panel height is chosen to facilitate comparison between panels].}
\label{fig_result_dimerobservable}
\end{figure}

\begin{figure}
    \includegraphics[width=0.45\textwidth]{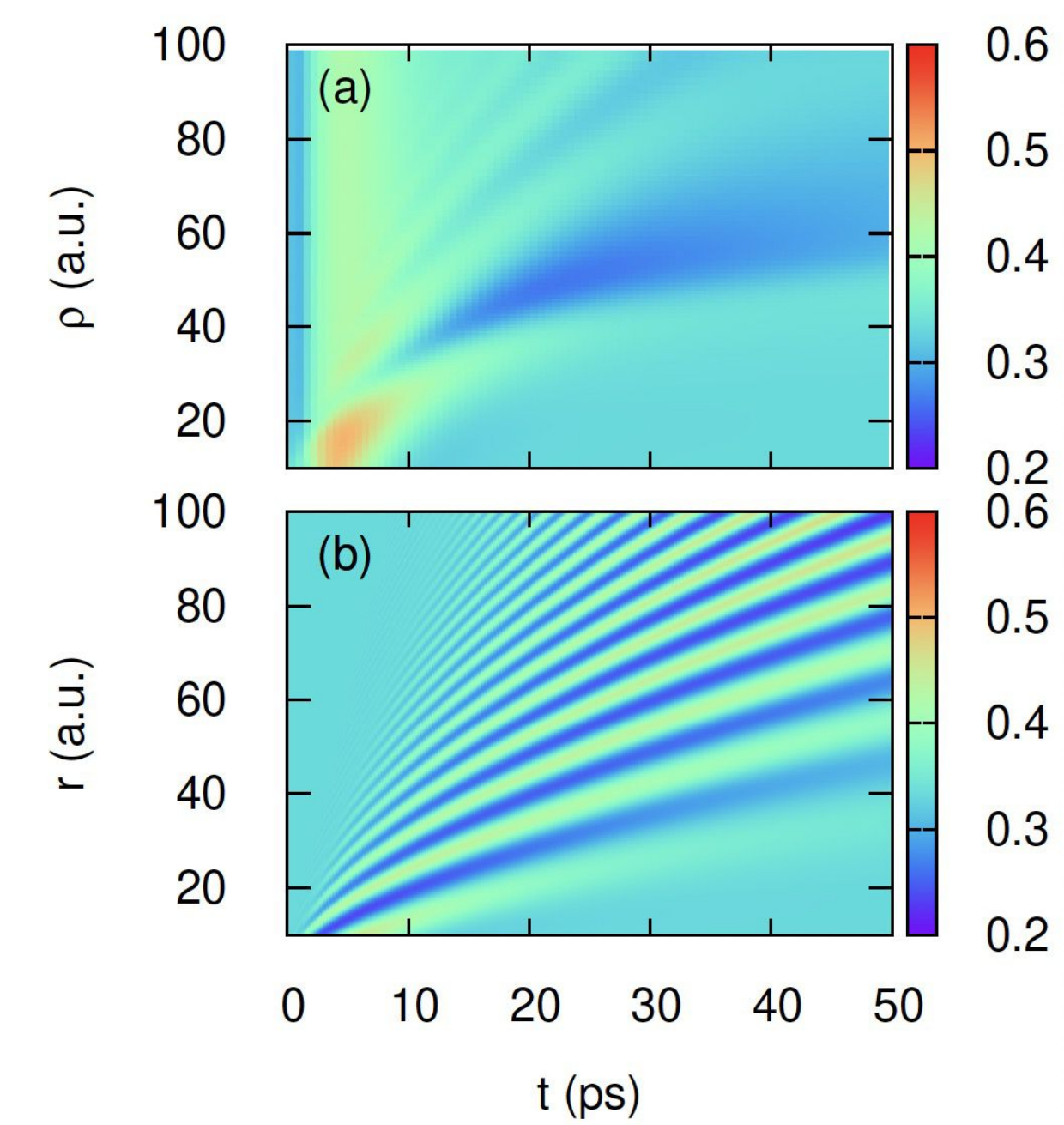}
    \caption{Trimer dynamics based on dimer input. 
    (a) Expectation value 
    $\langle \cos^2(\vartheta)\rangle(\rho,t)$ for the trimer.
    The trimer signal is determined using  the dimer dynamics as input; a scaling factor of $10,000$ is used  (see the main text and Appendix~\ref{appendix_dimer} for details). (b) Expectation value $\langle \cos^2(\vartheta)\rangle (r,t)$ for the helium dimer.  The legends on the right color code the expectation value. }
    \label{fig_result_dimermodel}
\end{figure}

Given the qualitative resemblance between the trimer and dimer dynamics (both display an interference when the alignment signal is plotted in a size-resolved manner, with the hyperradial and  distance coordinates taken as a proxy of the system size), it is instructive to make a more direct comparison. To this end, Figs.~\ref{fig_result_dimerobservable}(a)-\ref{fig_result_dimerobservable}(c) show the quantities $\langle \cos^2(\vartheta_{\text{ave}})\rangle (r,t)$,  $\langle \cos^2(\vartheta_{\text{max}})\rangle (r,t)$, and $\langle \cos^2(\vartheta_{\text{min}})\rangle (r,t)$, respectively. While the average and maximum values show a distinct finger-like pattern, the minimum values show a comparatively faint finger-like pattern. Since the minimum distance corresponds, in most cases, to a bound dimer, it seems reasonable that the interference in this observable is less pronounced than that in the other two observables.
For comparison, Fig.~\ref{fig_result_dimermodel}(b) shows the same observable for the dimer using the same $\delta$-function laser pulse and the same laser intensity and FWHM.
The resemblance of Fig.~\ref{fig_result_dimerobservable} and Fig.~\ref{fig_result_dimermodel}(b) is intriguing. 

The discussion in the previous paragraph motivates an approach that  constructs the time-dependent trimer wave packet based on only the dimer wave packet. While such a model may not  quantitatively capture the full trimer dynamics, it might provide insights into how much the third helium atom ``disturbs'' the time evolution of the other two. Appendix~\ref{appendix_dimer} develops a model that uses the dimer wave packet as input. With one free parameter, namely an overall scaling factor, the model produces the trimer observable shown in Fig.~\ref{fig_result_dimermodel}(a). The trimer dynamics constructed  from the ``dimer model''  shares several key features with the results obtained from the  full trimer wave packet. Specifically,  Fig.~\ref{fig_result_dimermodel}(a) shows a finger-like pattern and the number of ``fingers'' are about the same as what is seen in the full wave packet results.
Even though the dimer model captures the overall behavior, it should be noted that there exist differences, notably the more pronounced local maximum in the  $\rho \lesssim 20$~a.u. and $t \lesssim 10$~ps region and the more pronounced local minimum at $t \gtrsim 20$~ps and $\rho\approx 40-80$~a.u. [blue region in Fig.~\ref{fig_result_dimermodel}(a)].

\begin{figure}
    \includegraphics[width=0.45\textwidth]{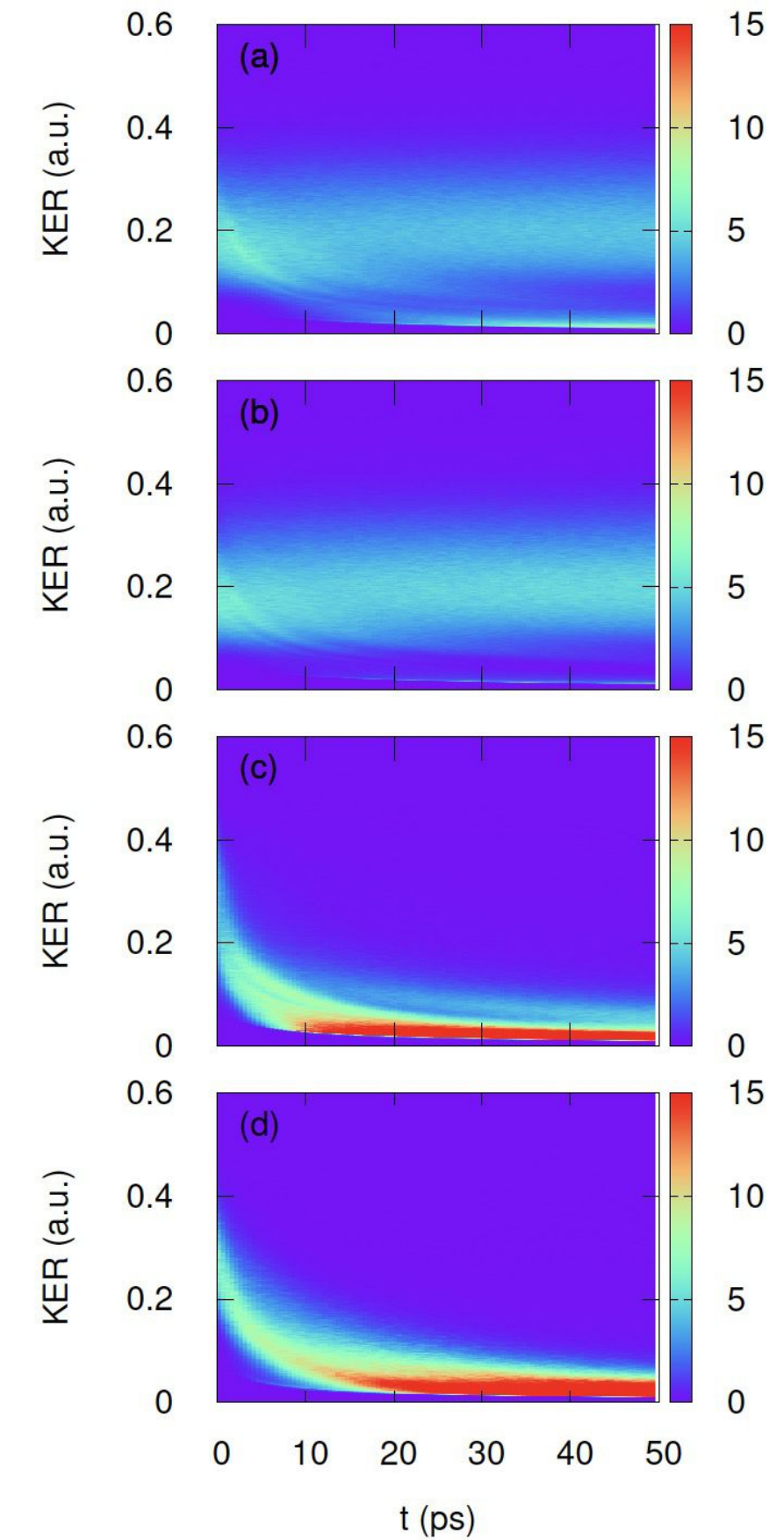}
     \vspace{-0.0cm}
    \caption{Analysis of kinetic energy release.
    (a) $P(\text{KER},t)$. 
    (b) $P^{(0)}(\text{KER},t)$.
    (c) $P^{(0,\text{scatt})}(\text{KER},t)$.
    (d) $P^{(2)}(\text{KER},t)$.
     The legends on the right color code the number of counts.
      The normalization is chosen such that the integral $\int_0^{\infty} P({z},t) d{z}=1$ for each time ($z$ is a placeholder for $\text{KER}$).}
    \label{fig_result_ker}
\end{figure}

From an experimental point of view, the quantity $P({\text{KER}},t)$ is a particularly important observable
since it is, unlike 
 the experimental determination of
    the probability distribution, 
     independent of the
    image reconstruction~\cite{doi:10.1126/science.aaa5601}; because of this, we refer to it as a ``direct''
    or ``unbiased'' observable.
Figure~\ref{fig_result_ker}(a) shows $P(\text{KER},t)$. Two contributions can be identified: an approximately time-independent contribution that is centered around $\text{KER}=0.2$~a.u. and a time-dependent contribution that moves to smaller KER with increasing time. To gain further insight, Figs.~\ref{fig_result_ker}(b), \ref{fig_result_ker}(c), and \ref{fig_result_ker}(d) show $P^{(0)}(\text{KER},t)$, $P^{(0,\text{scatt})}(\text{KER},t)$, and $P^{(2)}(\text{KER},t)$, respectively.
It can be seen that, as might be expected, the scattering part of the $J=0$ contribution and the $J=2$ contribution have appreciable contributions at roughly the same KER values at comparable times.
Interestingly, the scattered $J=0$ component contains one contribution that moves to smaller KER more slowly.

The results presented up to now were obtained for a delta-function pulse that corresponds to $\tau=311$~fs and $I=3.5 \times 10^{14}$~W/cm$^2$.
To provide some guidance on how the results change if the area under the peak is changed, Table~\ref{table_initial3} shows the probability $|c_{\text{trimer}}^{(\text{ground})}|^2$ to be in the trimer ground state at time $t=0^+$ for the same pulse length as used before but for four smaller intensities. 
Over the intensity range considered (negligible excitations of electronic degrees of freedom), the dependence of the ground state population at $t=0^+$ on $I$ is fit well by
$|c_{\text{trimer}}^{(\text{ground})}|^2=1-0.0153/(10^{14}\text{W/cm}^2) I -0.0127 /(10^{14}\text{W/cm}^2)^2 I^2$. The terms proportional to $I^0$, $I^1$, and $I^2$ can be understood by  Taylor-expanding the term $\exp[\imath \overline{\varphi}(\rho,\theta,\phi,\beta,\gamma,0^+)]$.
As the intensity is decreased, for the same pulse length, the experimental observation of the interference displayed in Fig.~\ref{fig_result_dynamics_external} requires averaging over more samples since the fraction of the wave packet that is unbound decreases. Even though a change of the pulse parameters is expected to lead to small changes in the observables, the overall behavior is expected to be robust to modest changes of the pulse length and intensity.
A change of the polarization vector, in contrast, would introduce appreciable changes.

\begin{table}
    \begin{tabular}{c|c|c|c|c}
      $I$ ($10^{14}$~W/cm$^2$) & 1.5 & 2.0 & 2.5 & 3.0 \\
     \hline $|c_{\text{trimer}}^{\text{(ground)}}|^2$ & 0.9497 & 0.9183 & 0.8815 & 0.8410 
        \end{tabular}
       \caption{Trimer ground state contribution $|c_{\text{trimer}}^{(\text{ground})}|^2$ to the wave packet at $t=0^+$ for a delta-function pulse corresponding to  $\tau=331$~fs and varying intensity $I$. The probabilities are calculated through direct integration.}
        \label{table_initial3}
        \end{table}

\section{Conclusion}
\label{sec_conclusion}
Historically, the determination of the eigen energies and eigen states of quantum mechanical three-body systems has played a pivotal role across various physics disciplines, ranging from high-energy physics (e.g., hyperons) to nuclear physics [e.g., the triton or carbon nucleus (treated as consisting of three $\alpha$ particles)] to condensed matter physics (e.g., trion) to chemical physics (e.g., H$_2$O or HCN molecules) to atomic physics [e.g., the helium atom (nucleus plus two electrons) or light van der Waals trimers]~\cite{Naidon_2017,BRAATEN2006259,greene-rev-mod-physics}. This paper considered the dynamics
of the van der Waals helium trimer, focusing on the low-energy regime where each atom can be treated as a structureless point particle. The dynamics was initiated through a short intense laser pulse that imparted angular momentum onto the helium trimer, which was assumed to be in its ground state prior to the application of the laser pulse. The ensuing external-field induced three-body dynamics is interesting since---among other things---the trimer starts off in a state that has universal characteristics, i.e., properties that are approximately, or at least in part, set by the $s$-wave scattering length. Since the three-particle dynamics also plays a pivotal role in condensed matter physics, where lasers are employed as probes, and in ultracold atomic gases, where magnetic field quenches are used to induce dynamics, the techniques developed and findings reported in the present paper are expected to be relevant to a broad audience.

The trimer has one more particle than the corresponding dimer and may be viewed as a step toward building up the bulk system. While, arguably, a three-body system might generally be considered much closer to the small-particle limit than the large-particle limit, several  studies show that the three-body system  exhibits precursors of what is being observed in larger systems, at least for selected observables (see, e.g., Refs.~\cite{cui-threebody,levinsen}). Adopting a bottom-up perspective, the trimer can be viewed as the simplest non-trivial system, which allows one to quantify the impact of ``spectator'' particles.  Considering static properties, 
we know  that adding a third atom to the helium dimer has an enormous effect: The binding energy increases by a factor of nearly 100 and the size decreases correspondingly by a factor of nearly 10.
This work explored what happens in the time domain. 
Employing an additive laser-molecule interaction potential,
we obtained the
first theory-based helium trimer movie; the movie shows intricate coupling of the vibrational and rotational degrees of freedom.
We showed that the alignment signal $\langle \cos^2 (\vartheta) \rangle(r,t)$ of the helium trimer can be determined approximately using the dimer wave packet as input, provided one allows for one adjustable scaling parameter. This suggests that the laser acts predominantly on one two-body bond, causing the third atom to adjust in response to changes of the first and second atoms.

This paper analyzed the dissociative dynamics of the helium trimer over a time of $50$~ps. This dynamics can be observed experimentally in state-of-the-art pump-probe spectroscopy setups. The 
 distribution of the kinetic energy release, 
 which can be measured directly experimentally, i.e.,  without the need of applying reconstruction algorithms, was analyzed. Specifically, the contributions due to different wave packet components as well as bound versus scattering components were identified. The pulse-induced dynamics of the orientational degrees of freedom (Euler angles $\beta$ and $\gamma$) was analyzed. While 
the average of these angles does not display appreciable dynamics, their hyperradius- or distance-resolved generalizations do display appreciable dynamics, including a finger-like pattern that was identified as a fingerprint of an interference of two relative orbital angular momentum channels, namely the  $J=0$ and $J=2$ channels.

The numerical approach developed in this work can be applied to other systems. While the number of channels needed for heavier and more deeply-bound trimers might be prohibitively  large, the approach should be  suitable for describing other weakly-bound trimers. Explicit treatment of the laser-molecule interaction, which was in the present work modeled as a $\delta$-function  and not as a Gaussian pulse, should be feasible. The successful prediction of a subset of the trimer dynamics based on only two-body input suggests that generalizations of this model might provide useful guidance for the dynamics of larger weakly-bound van der Waals clusters.
Finally, it should be noted that a detailed side-by-side comparison of a subset of  the presented theory results and experimental data is currently being pursued~\cite{our-experiment-theory-paper}.

  \section{Acknowledgement}
  D.B. acknowledges support by the National Science Foundation through grant numbers PHY-2110158 and PHY-2409311. Q.G. acknowledges support by the National Science Foundation through grant number PHY-2409600 and support by Washington State University through the Claire May \& William Band Distinguished Professorship Award and the New Faculty Seed Grant. This work used the OU Supercomputing Center for Education and Research (OSCER) at the University of Oklahoma (OU) and the Kamiak High Performance Computing Cluster at Washington State University. This research was supported in part by grant NSF PHY-1748958 to the Kavli Institute for Theoretical Physics (KITP).
   
\appendix

\section{Rigid-body dynamics}
\label{appendix_rigid}
This appendix discusses the trimer dynamics under the rigid-body assumption, i.e., under the assumption that the laser changes the orientation of the trimer, which is governed by the angles $\beta$ and $\gamma$, but not its  shape. Dropping terms that contain derivatives with respect to  $\rho$, $\theta$, and $\phi$, 
 the rigid-body kinetic energy operator takes the form 
\begin{eqnarray}
\label{eq_kinetic}
    T_{\text{rigid}} = -\frac{\hbar^2}{ M_{\text{hyper}} \rho^2 \sin^2(2 \theta)} \times \nonumber \\
    \bigg[
    -{J}_{\text{op}}^2 + {J}_{\text{op},z}^2 - \frac{\cos(2 \theta)}{2} \left( {J}_{\text{op},+}^2 + {J}_{\text{op},-}^2 \right) - \nonumber \\ \frac{\tan^2(2 \theta)}{2} {J}_{\text{op},z}^2\bigg],
    \nonumber \\
\end{eqnarray}
where the angular momentum operators
$J_{\text{op}}$,
$J_{\text{op},z}$,
$J_{\text{op},+}$, and
$J_{\text{op},-}$
have the following properties:
\begin{eqnarray}
    J_{\text{op}}^2 D_{M',M}^{(J)}(\alpha,\beta,\gamma)=
    J(J+1) D_{M',M}^{(J)}(\alpha,\beta,\gamma),
\end{eqnarray}
\begin{eqnarray}
    J_{\text{op},z}^2 D_{M',M}^{(J)}(\alpha,\beta,\gamma)=
    M^2 D_{M',M}^{(J)}(\alpha,\beta,\gamma),
\end{eqnarray}
\begin{eqnarray}
    J_{\text{op},+}^2 D_{M',M}^{(J)}(\alpha,\beta,\gamma)=
    A_{J,M}^+ D_{M',M+2}^{(J)}(\alpha,\beta,\gamma),
\end{eqnarray}
and
\begin{eqnarray}
    J_{\text{op},-}^2 D_{M',M}^{(J)}(\alpha,\beta,\gamma)=
    A_{J,M}^- D_{M',M-2}^{(J)}(\alpha,\beta,\gamma).
\end{eqnarray}
In Eq.~(\ref{eq_kinetic}), $\rho$ and $\theta$ are considered parameters, whose values are determined by the shape of the trimer. 
The quantities $A_{J,M}^{\pm}$ are defined in Eq.~(\ref{eq_aplusminus}).

Under the rigid-body assumption, the interaction potential $V_{\text{trimer}}$ reduces to a constant that can be dropped as it simply introduces a trivial overall phase into the time dependent wave packet. The laser-trimer interaction depends on the internuclear distances $r_{12}$, $r_{13}$, and $r_{23}$, which are constant, as well as the angles $\vartheta_{12}$, $\vartheta_{13}$, and $\vartheta_{23}$, which can be expressed as
\begin{eqnarray}
    \cos \vartheta_{12}=\frac{\overline{d} \rho}{r_{12}}
    \cos \phi \cos \theta \sin \beta 
    \cos \gamma 
    + \nonumber \\
    \frac{\overline{d} \rho}{r_{12}} \sin \phi \sin \theta \sin \beta \sin \gamma,
\end{eqnarray}
\begin{eqnarray}
    \cos \vartheta_{13}=\frac{\overline{d} \rho}{2r_{13}}
    \cos \theta \left(\cos \phi + \sqrt{3} \sin \phi \right) \sin \beta 
    \cos \gamma 
    + \nonumber \\
    \frac{\overline{d} \rho}{2r_{13}} \sin \theta \left(-\sqrt{3}\cos \phi + \sin \phi\right)  \sin \beta \sin \gamma,\nonumber \\
\end{eqnarray}
and
\begin{eqnarray}
    \cos \vartheta_{23}=-\frac{\overline{d} \rho}{2r_{23}}
    \cos \theta \left(\cos \phi - \sqrt{3} \sin \phi \right) \sin \beta 
    \cos \gamma 
    - \nonumber \\
    \frac{\overline{d} \rho}{2r_{23}} \sin \theta \left(\sqrt{3}\cos \phi + \sin \phi\right)  \sin \beta \sin \gamma, \nonumber \\
\end{eqnarray}
where $\overline{d}=\sqrt{2}/3^{1/4}$.
The expressions are arranged such that the Euler angles are included at the end of each term. It can be seen that the laser-trimer potential depends on the combinations 
$\sin \beta \cos \gamma$ and $ \sin \beta \sin \gamma$,
which can be expressed in terms of the Wigner-D matrices,
\begin{eqnarray}
    \sin \beta \sin \gamma = \nonumber \\ \frac{\imath}{\sqrt{2}} \left[ D_{0,-1}^{(1)}(\alpha,\beta,\gamma)+
    D_{0,1}^{(1)}(\alpha,\beta,\gamma) \right]
\end{eqnarray}
and
\begin{eqnarray}
    \sin \beta \cos \gamma = \nonumber \\
    \frac{1}{\sqrt{2}} \left[ D_{0,-1}^{(1)}(\alpha,\beta,\gamma)-
    D_{0,1}^{(1)}(\alpha,\beta,\gamma) \right].
\end{eqnarray}
The preceding expressions show explicitly that the laser-trimer interaction model is independent of $\alpha$.
Using that the laser-trimer potential depends on $\cos ^2 \vartheta_{jk}$ and not on $\cos  \vartheta_{jk}$,
it can be verified that the laser-trimer interaction model leads, provided one starts prior to the application of the laser pulse in a state with $J=0$ and $M=0$, to the occupation of wave packet components with only  even $J$ (and $|M|=0,\pm2,\cdots$).

Assuming a specific internal trimer structure (i.e., fixing the values of $\rho$, $\theta$, and $\phi$ in, e.g., an equilateral triangle or linear structure) and starting in the state $D_{0,0}^{(0)}(\alpha,\beta,\gamma)$, the time evolution of the wave packet 
$\Psi_{\text{rigid}}(\beta,\gamma,t)$ can be written as
\begin{eqnarray}
    \Psi_{\text{rigid}}(\beta,\gamma,t)=
    \nonumber \\
    \sum_{J=0,2,\cdots} \sum_{M=0,\pm 2,\cdots}
    c_{0,M}^{(J)}(t)
    D_{0,M}^{(J)}(\alpha,\beta,\gamma).
\end{eqnarray}
The solution to the time-dependent Schr\"odinger equation can thus be found by
constructing the rigid-body Hamiltonian matrix $H_{\text{rigid}}$ in the $D_{0,M}^{(J)}$ basis and determining 
the time dependence of the expansion coefficients $c_{0,M}^{(J)}$ for the initial condition $c_{0,M}^{(J)}(t=0)=\delta_{J,0}$.
The Hamiltonian matrix contains finite off-diagonal elements due to the Coriolis coupling terms that exist even in the absence of the external laser pulse as well as due to the laser-trimer interaction potential. 

 By construction, the rigid-body model does not provide any insights into observables that depend on the internal coordinates $\rho$, $\theta$,
and $\phi$.
Lines in Figs.~\ref{fig_result_average_euler}(a) and \ref{fig_result_average_euler}(b) show the expectation values $\langle \cos^2(\beta) \rangle(t)$ and $\langle \cos^2(\gamma) \rangle(t)$, respectively, as a function of time,
 using the same  $\delta$-function pulse and parameters as those employed in the full calculations (see main text).
The red dashed and blue solid lines are for two different internal structures, namely an approximately equilateral triangle 
with 
$r_{12}=7.833$~a.u. and  $r_{13}=r_{23}=7.478$~a.u., 
and an approximately linear molecule with  
$r_{12}=r_{13}=8.643$~a.u. and $r_{23}=4.879$~a.u..
The oscillatory behavior of the rigid-body model can be understood from the field-free rigid-body eigen spectrum: a change of the structure leads to a change of the moments of inertia of the rigid body~\cite{Nielsen_1998}.

\section{Modeling trimer dynamics using dimer dynamics input }
\label{appendix_dimer}

This appendix develops an approximate treatment that
results in an expression for the time-dependent
trimer wave packet in terms of the time-dependent dimer wave packet that is obtained by exposing the ground state helium dimer
to the same laser  as the ground state helium trimer.
It is shown in Sec.~\ref{sec_results} that this approximate treatment
captures several characteristics of the full trimer dynamics
qualitatively.

To establish a reference point, we consider the dynamics of the helium
dimer that ensues in response to a $\delta$-function laser pulse.
We write the time-dependent dimer wavepacket $\Psi_{\text{dimer}}(r_{12},\vartheta_{12},t)$
as
\begin{eqnarray}
  \label{eq_ansatz_dimer_dimer}
  \Psi_{\text{dimer}}(r_{12},\vartheta_{12},t)=
  \left[1+
  g_{\text{dimer}}(r_{12},\vartheta_{12},t) \right] \times  \nonumber \\
  \exp \left( - \frac{\imath E_{\text{dimer}}^{(\text{ground})} t}{\hbar} \right)
  \psi_{\text{dimer}}^{(\text{ground})}(r_{12}),
\end{eqnarray}
where $E_{\text{dimer}}^{(\text{ground})}$ and
$\psi_{\text{dimer}}^{(\text{ground})}(r_{12})$ denote the eigen energy and
eigen state of the ground state dimer.
Inserting the ansatz (\ref{eq_ansatz_dimer_dimer})
into the time-dependent Schr\"odinger equation for $t>0^+$,
we find that the unknown function
$g_{\text{dimer}}(r_{12},\vartheta_{12},t)$
obeys the equation
\begin{eqnarray}
  \label{eq_dimer_effective}
  \imath \hbar \frac{\partial g_{\text{dimer}}(r_{12},\vartheta_{12},t)}
         {\partial t}
         =
         -\frac{\hbar^2}{2\overline{\mu}} 
         \frac{\partial ^2 g_{\text{dimer}}(r_{12},\vartheta_{12},t)}{\partial \vec{r}_{12}^2}
           -
           \nonumber \\
           \frac{\hbar^2}{\overline{\mu} \psi_{\text{dimer}}^{(\text{ground})}(r_{12})}
\frac{\partial \psi_{\text{dimer}}^{(\text{ground})}(r_{12})}{\partial r_{12}}
           \frac{\partial g_{\text{dimer}}(r_{12},\vartheta_{12},t)}{\partial r_{12}},\nonumber \\
\end{eqnarray}
where $\overline{\mu}$ denotes the reduced mass of the dimer, $\overline{\mu}=\mu/2$.
The logarithmic derivative of $\psi_{\text{dimer}}^{(\text{ground})}(r_{12})$
in the
second term on the right hand side of Eq.~(\ref{eq_dimer_effective})
can be interpreted as an effective force.
The initial condition is given by
\begin{eqnarray}
\label{eq_initial_dimer}
  1+g_{\text{dimer}}(r_{12},\vartheta_{12},0^+)
      = \nonumber \\
      \exp \left[
    -\imath C V_{\text{laser-dimer}}(r_{12},\vartheta_{12})
    \right] .
\end{eqnarray}
If the laser pulse is sufficiently weak, we can Taylor expand
the exponential. This yields
\begin{eqnarray}
  \label{eq_initial_dimer_mod}
  g_{\text{dimer}}(r_{12},\vartheta_{12},0^+)=- \imath C V_{\text{laser-dimer}}(r_{12},\vartheta_{12}).
\end{eqnarray}
In what follows, we develop an approximate description  that
reduces the trimer dynamics to equations that are formally equivalent to
Eqs.~(\ref{eq_dimer_effective})
and (\ref{eq_initial_dimer_mod}). Since $g_{\text{dimer}}(r_{12},\vartheta_{12},t)$
can be calculated using the approach discussed in Ref.~\cite{nature-physics}  [the corresponding alignment signal is shown in Fig.~\ref{fig_result_dimermodel}(b)], this implies that the approximate trimer dynamics can be constructed
using $g_{\text{dimer}}(r_{12},\vartheta_{12},t)$ as input.

For the trimer, we start by expressing the time-dependent
wave packet in terms of the unknown function
$g(\rho,\theta,\phi,\beta,\gamma,t)$,
\begin{eqnarray}
  \label{eq_ansatz_dimer}
  \Psi(\rho,\theta,\phi,\beta,\gamma,t)
  =
  g(\rho,\theta,\phi,\beta,\gamma,t) \times \nonumber \\
  \exp \left( - \frac{\imath E_{\text{trimer}}^{(\text{ground})} t}{\hbar} \right)
  \sqrt{\frac{3}{8 \pi^3}} \psi_{\text{trimer}}^{(\text{ground})}(\rho,\theta,\phi),
\end{eqnarray}
where
\begin{eqnarray}
  \label{eq_gfactor_delta}
  g(\rho,\theta,\phi,\beta,\gamma,0^+)= \exp
  [ \imath \overline{\varphi}(\rho,\theta,\phi,\beta,\gamma,0^+)].
\end{eqnarray}
In writing Eq.~(\ref{eq_gfactor_delta}), we modeled---as before---the laser
pulse by a $\delta$-function in time
[specifically, $\overline{\varphi}(\rho,\theta,\phi,\beta,\gamma,0^+)$ is defined in Eq.~(\ref{eq_phase2})].
Inserting Eq.~(\ref{eq_gfactor_delta}) into the time-dependent Schr\"odinger
equation for $t>0^+$, we obtain 
\begin{eqnarray}
  \imath \hbar \frac{\partial g(\rho,\theta,\phi,\beta,\gamma,t)}{\partial t}
  = \nonumber \\
  \sum_{j=1}^3
  \left[
    -\frac{\hbar^2}{2\mu} 
    \frac{\partial^2}{\partial \vec{r}_j^2}
+
\vec{F}_j(\rho,\theta,\phi)
        \cdot 
        \frac{\partial}{\partial \vec{r}_j
        } 
        \right]  
  g(\rho,\theta,\phi,\beta,\gamma,t),\nonumber \\
\end{eqnarray}
where the time-independent
effective force $\vec{F}_j(\rho,\theta,\phi)$ is given by
\begin{eqnarray}
  \vec{F}_j(\rho,\theta,\phi) =-
  \frac{\hbar^2}{\mu \psi_{\text{trimer}}^{(\text{ground})}(\rho,\theta,\phi) }
    \frac{\partial \psi_{\text{trimer}}^{(\text{ground})}(\rho,\theta,\phi)}{\partial \vec{r}_j
    } .\nonumber \\
  \end{eqnarray}
To proceed, we assume that the laser pulse is perturbative. This motivates us
to Taylor
expand the initial state,
\begin{eqnarray}
  g(\rho,\theta,\phi,\beta,\gamma,0^+)= \nonumber \\
  1 - \imath C \sum_{j=1}^2 \sum_{k=j+1}^3 V_{\text{laser-dimer}}(r_{jk},\vartheta_{jk}),
\end{eqnarray}
and to constrain the functional form of
$g(\rho,\theta,\phi,\beta,\gamma,t)$,
\begin{eqnarray}
  g(\rho,\theta,\phi,\beta,\gamma,t)=1+
  \sum_{j=1}^2 \sum_{k=j+1}^3 g_{2}(r_{jk},\vartheta_{jk},t).
\end{eqnarray}
 As discussed in the context of Fig.~\ref{fig_phase}, the assumption that the magnitude of the phase $\overline{\varphi}(\rho,\theta,\phi,\beta,\gamma,0^+)$ is small compared to one holds, for the laser parameters considered in this work, for a large portion of the configuration space but not everywhere. In light of this, the model developed in this appendix is expected to provide a qualitatively but not quantitatively correct description.  
Under 
 the stated assumptions, we have
\begin{eqnarray}
  g_{2}(r_{jk},\vartheta_{jk},0^+)=- \imath C V_{\text{laser-dimer}}(r_{jk},\vartheta_{jk}).
\end{eqnarray}
It can be seen that
\begin{eqnarray}
  \label{eq_equivalence}
  g_{2}(r_{jk},\vartheta_{jk},0^+)=g_{\text{dimer}}(r_{jk},\vartheta_{jk},0^+),
\end{eqnarray}
i.e.,
the initial conditions of the two-body components $g_{2}(r_{jk},\vartheta_{jk},0^+)$
of the trimer wave packet are the same as the initial condition for the
function $g_{\text{dimer}}(r_{12},\vartheta_{12},t)$ that enters into the dimer wave packet.

In addition, we approximate the
trimer ground state $\psi_{\text{trimer}}^{(\text{ground})}(\rho,\theta,\phi)$
by  a variational Jastrow ansatz,
  \begin{eqnarray}
    \psi_{\text{trimer}}^{(\text{ground})}(\rho,\theta,\phi)
    =
    \prod_{j=1}^2 \prod_{k>j}^3
    f(r_{jk});
  \end{eqnarray}
    this commonly employed ansatz has
  been shown to
  reproduce 
   $95$~\% of the full ground state energy of the helium trimer~\cite{AJ}.
  Neglecting several derivatives that originate in the force
  term, the function $g_2(r_{jk},\vartheta_{jk},t)$
  is found to obey the Schr\"odinger like equation,
  \begin{eqnarray}
    \label{eq_trimer_model}
    \imath \hbar
    \frac{\partial g_{2}(r_{jk},\vartheta_{jk},t)}{\partial t}
    =
      -\frac{\hbar^2}{2 \overline{\mu}} 
      \frac{\partial^2 g_{2}(r_{jk},\vartheta_{jk},t)}{\partial \vec{r}_{jk}^2} 
      -
      \nonumber \\
      \frac{\hbar^2}{\overline{\mu}}
      \frac{1}{f(r_{jk})} \frac{\partial f(r_{jk})}{\partial r_{jk}}
      \frac{\partial g_{2}(r_{jk},\vartheta_{jk},t)}
           {\partial r_{jk}} .
  \end{eqnarray}
  Comparison of Eqs.~(\ref{eq_trimer_model})
  and (\ref{eq_dimer_effective})
  shows that---keeping Eq.~(\ref{eq_equivalence}) in mind---the time evolution of
  $g_{2}(r_{jk},\vartheta_{jk},t)$ is equivalent to that of
  $g_{\text{dimer}}(r_{12},\vartheta_{12},t)$ provided
  $f(r_{jk})$
  is chosen to be equal to
  $\psi_{\text{dimer}}^{\text{(ground)}}(r_{jk})$.
  While the resulting Jastrow
  wave function
  does provide a rather poor description of the trimer
  ground state wave function (the trimer is much smaller than the dimer; see Fig.~\ref{fig_pair}),
  the resulting effective force is expected to have roughly
  the correct functional form but a magnitude that is notably smaller than a more accurate
  treatment would yield. The reason is that the short-distance behavior of the pair distribution
  functions of the dimer and trimer is, except for a scaling factor that is related to Tan's two-body contact, very similar~\cite{TAN20082987,TAN20082971,TAN20082952,AJ,PhysRevA.101.010501}. Since the laser has the largest effect at short distances,
  setting $f(r_{jk})=\psi_{\text{dimer}}^{\text{(ground)}}(r_{jk})$
  may thus provide a reasonable description of the trimer dynamics.

  The physical picture behind the model developed in this section is that the dynamics
  of each pair of atoms
  within the trimer follows dynamics analogous to the dynamics an isolated
  atom pair would follow.
The results of the model developed in this appendix are shown in Fig.~\ref{fig_result_dimermodel}(a).


\begin{thebibliography}{98}
\bibitem{footnote_kick}
  It should be noted that the results only depend on the area under the Gaussian pulse. This implies that our results also apply to pulse lengths and intensities for which $\tau\rightarrow a \tau$ and $I \rightarrow a^{-1} I$, where $a$ is a real number. The laser intensity $I$ is related to $\overline{\epsilon}$ via $I=\epsilon_0 c \overline{\epsilon}^2/2$, where $\epsilon_0$ and $c$ denote the vacuum permittivity and the speed of light, respectively.
\bibitem{footnote_conversion}
The time scale is estimated using $(\text{time scale})=2 \pi \hbar/ (\text{energy scale})$.
\end{thebibliography}
\end{document}